\documentclass[letterpaper,twocolumn,10pt,anonymous]{article}
\usepackage{usenix2019_v3}

\usepackage{tikz}
\usepackage{amsmath}

\usepackage{filecontents}

\usepackage{xcolor}
\usepackage{graphicx}
\usepackage{booktabs}
\usepackage{amsmath}
\usepackage{subcaption}
\usepackage{adjustbox}
\usepackage{multirow}
\usepackage{siunitx}

\usepackage{enumitem} 
\usepackage{here}
\usepackage{xurl}
\usepackage[]{hyperref} 
\hypersetup{
    colorlinks=true,urlcolor=purple,linkcolor=cyan,citecolor=cyan
}
\usepackage[labelformat=simple]{subcaption}

\usepackage[available]{usenixbadges}

\newif\ifdraft
\draftfalse 

\ifdraft
    \newcommand{\ttsuchiy}[1]{\textcolor{blue}{[[Taro: #1]]}}
    \newcommand{\nicolasc}[1]{\textcolor{brown}{[[Nicolas: #1]]}}
    \newcommand{\jindongd}[1]{\textcolor{red}{[[Mark: #1]]}}
    \newcommand{\ksoska}[1]{\textcolor{purple}{[[Kyle: #1]]}}

\else
    \newcommand{\ttsuchiy}[1]{}
    \newcommand{\nicolasc}[1]{}
    \newcommand{\jindongd}[1]{}
    \newcommand{\ksoska}[1]{}
\fi

\begin{document}

\date{}

\title{Blockchain Address Poisoning}

\author{
{\rm Taro Tsuchiya\textsuperscript{\rm 1}},
{\rm Jin-Dong Dong\textsuperscript{\rm 1}},
{\rm Kyle Soska\textsuperscript{\rm 2}},
{\rm Nicolas Christin\textsuperscript{\rm 1}} \\
\textsuperscript{\rm 1}Carnegie Mellon University,
\textsuperscript{\rm 2}Independent
}

\maketitle

\begin{abstract}
In many blockchains, e.g., Ethereum, Binance Smart Chain (BSC),
the primary representation used for wallet addresses is a hardly
memorable 40-digit hexadecimal string. As a result, users often
select addresses from their recent transaction history, which enables
\textit{blockchain address poisoning}. The adversary first generates
lookalike addresses similar to one with which the victim has previously
interacted, and then engages with the victim to ``poison'' their
transaction history. The goal is to have the victim mistakenly send
tokens to the lookalike address, as opposed to the intended recipient.
Compared to contemporary studies, this paper provides four notable
contributions. First, we develop a detection system and perform
measurements over two years on both Ethereum and BSC. We identify 13~times
more attack attempts than reported previously---totaling 270M on-chain
attacks targeting 17M victims. 6,633 incidents have caused at least
83.8M USD in losses, which makes blockchain address poisoning one of
the largest cryptocurrency phishing schemes observed in the wild.
Second, we analyze a few large attack entities using improved clustering
techniques, and model attacker profitability and competition. Third,
we reveal attack strategies---targeted populations, success conditions
(address similarity, timing), and cross-chain attacks. Fourth, we
mathematically define and simulate the lookalike address generation
process across various software- and hardware-based implementations, and
identify a large-scale attacker group that appears to use GPUs. We also
discuss defensive countermeasures. 
\end{abstract}

\section{Introduction}
\label{sec:intro}

\begin{figure*}[]
  \begin{center}
    \begin{subfigure}{0.3\linewidth}
      \includegraphics[width=\textwidth]{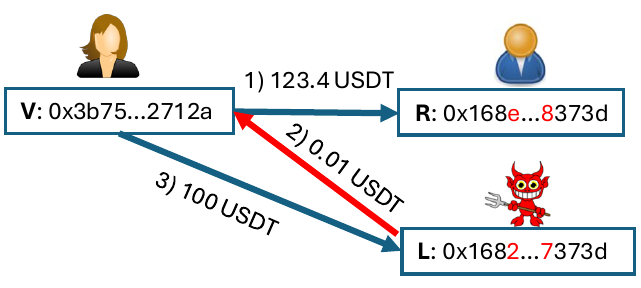}
      \caption{\footnotesize Tiny transfer}
      \label{subfig:tiny_transfer}
    \end{subfigure}
    \begin{subfigure}{0.3\linewidth}
     \includegraphics[width=\textwidth]{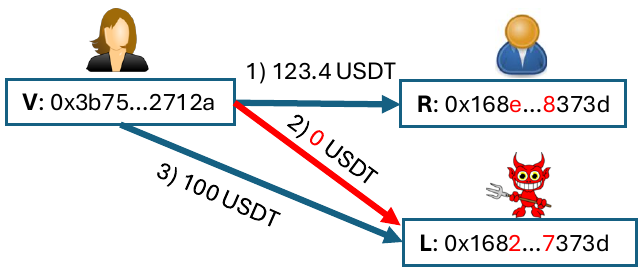}
    \caption{\footnotesize Zero-value transfer}
    \label{subfig:zero_transfer}
    \end{subfigure}
    \begin{subfigure}{0.3\linewidth}
     \includegraphics[width=\textwidth]{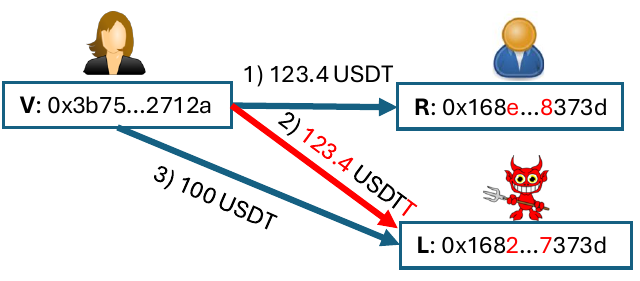}
    \caption{\footnotesize Counterfeit token transfer}
    \label{subfig:fake_transfer}
    \end{subfigure}
    \caption{Three types of poisoning transfers}
    \label{fig:attack_sketch}
  \end{center}
\end{figure*}

Most modern blockchains rely on ``wallets'' for monetary transfers. 
Wallet addresses are derived from cryptographic public keys and are often represented by long, hard to memorize, strings.
However, most currency or token transfers require users to manually input or select the recipient's wallet address. 
A common practice is to copy and paste addresses or select an address with which one has previously interacted. This introduces a new attack, \emph{blockchain address poisoning}.  

An adversary generates one or more ``lookalike'' address(es) whose first and last characters match those of an address the victim often interacts with. 
The attacker then floods (``poisons'') the target's transaction history with the lookalike address(es).
As a result, the victim may erroneously send digital assets to a lookalike address instead of the intended address.
In contrast to traditional bank transfers, blockchain transactions are irreversible, making fund recovery challenging, and thereby exacerbating the severity of any mistakes.
In this paper, we attempt to characterize the economics of address poisoning and to uncover attackers' strategies and capabilities through large-scale measurements and simulations. 

We design a detection system and run it between July 1, 2022, and June 30, 2024, on Ethereum and Binance Smart Chain (BSC).
Combining both chains, we identify over 270M attack attempts (i.e., 13 times more than the previous efforts: 21M~\cite{ye2024interface} and 14M~\cite{guan2024characterizing}) that target over 17M victims. 
Out of those, attackers successfully received 6,633 transfers, adding up to over 83.8M USD in ill-gotten gains.
This attack appears to be one of the largest cryptocurrency phishing schemes observed in the wild. 
We evaluate our detector by 1) the external dataset (100\% precision, 97.2\% recall), 2) analyzing its false negatives, 3) manually verifying the phished cases, 4) checking the smart contracts used for those attacks, and 5) comparing detection outcomes with different parameters (e.g., window size and similarity threshold).


Attackers typically optimize their operations by bundling several transfers into a single transaction and reusing addresses and contracts.
This behavior allows us to link attack transfers through the ``guilt-by-association'' heuristic~\cite{xia2021trade}.
After removing transactions from unrelated bots that occasionally copy poisoning transactions from attackers, we can establish the existence of a few large distinct attack groups which we use to model the economics of address poisoning.
For each group, we calculate revenues (i.e., a sum of phished amounts) and costs (i.e., total attack transaction costs).
Despite high variability, 
most large groups manage to stay profitable by succeeding often, while smaller groups sometimes suffer from loss (by losing to the competition). 

We also analyze attack strategies. 
First, we evaluate the characteristics of the targeted victims to learn what attackers are looking for. 
We find that targets are more likely to have had higher balances at the time of the attack, conducted more transactions, and transferred a larger amount compared to random stablecoin users. 
Second, we investigate what influences attack success. 
We find that victims are more likely to choose a lookalike address that is more similar to or appears earlier than others.   
Third, given that addresses are compatible across Ethereum and BSC, some attack groups re-use lookalike addresses and target victims across chains. 

To infer attackers' (hardware) lookalike address generation capabilities, we 
simulate address generation rates.
We implement our scripts both naively and with optimizations 
and compare the performance over different CPUs and GPUs.
Based on our benchmarks, 
one sophisticated attack group appears to be using GPUs 
while others seem to rely on CPUs.

Finally,  
we propose a few mitigations for address poisoning attacks at the protocol, contract, wallet, and user levels.

Compared to contemporary studies~\cite{ye2024interface,guan2024characterizing}, our paper offers the following contributions.
First, our work detects attacks across multiple chains (BSC, Ethereum); in doing so, it identifies 13 times more attacks than prior work, and uncovers cross-chain attacks.
Second, we devise improved clustering techniques, and use them to model group profitability and competition.
Third, we perform novel experiments (e.g., victim profiling, success conditions, address generation modeling) to provide deeper insights into attack strategies and attacker computational capability.

\section{Background}
\label{sec:bg}
We next discuss 1) basic blockchain primitives (smart
contracts, ERC-20 tokens), 2) wallet types and addresses, and 3)
differences between address poisoning and 
phishing attacks.

\subsection{Blockchain primitives}
Most popular modern public blockchains are either UTXO (Unspent
Transaction Output)-based or account-based. On UTXO-based chains such
as Bitcoin, address reuse is minimal: most addresses are used only for
two transactions (one to receive some money, one to spend it). On the
other hand, account-based chains such as Ethereum see significant address reuse.
Account-based chains are therefore acutely susceptible to address poisoning as users
frequently reference recipient addresses from past transactions. In this
paper, we only focus on account-based chains, so we use ``account'' and
``wallet addresses'' interchangeably.

Ethereum (and derived blockchains) features two types of
addresses: wallet addresses (also called External Owned Accounts\textemdash EOA) and
contract addresses. Wallet addresses are to send Ether (ETH, the native currency) or interact with smart contracts (whose
addresses are contract addresses).
Smart contracts are executed by all nodes using the
Ethereum Virtual Machine (EVM). Smart contracts are developed in a
higher-level programming language (e.g., Solidity), and then compiled
to ``bytecode'' intelligible by the EVM before being deployed. Smart
contracts allow users to perform actions, such as token implementation, token transfer. Of particular interest, \emph{ERC-20 tokens}
(e.g., USDT, USDC) are fungible assets, defined by the ERC-20 framework
\cite{ef_eip20}, for which creators specify a token name, symbol, and value
unit. Essentially, ERC-20 tokens allow developers to implement other
cryptocurrencies or payment systems on top of Ethereum.

For every action (e.g., sending ETH or ERC-20 tokens, or more generally, interacting
with any contract), users pay ``gas,'' that is, transaction fees
dependent on the computational complexity of the action requested.
Other chains, such as Binance Smart Chain (BSC), that inherit Ethereum's
structure are called EVM-compatible chains. Notably, wallet addresses
are compatible across EVM chains, allowing users to hold tokens across
chains in one wallet.

\subsection{Wallet and blockchain address}
Most users will interface with the blockchain via wallet software such as MetaMask or Phantom.
These powerful tools enable users to have full control over their funds and perform complex tasks 
such as engaging in DeFi (decentralized finance) activities. 
When sending direct payments, users must often manually specify the recipient address,
exacerbating usability challenges---since addresses are represented
by long (40-character), hardly memorable, hexadecimal strings. 
This UX challenge facilitates a significant threat from phishing or scams and motivates several concerns about wallet interfaces (e.g., the
absence of strong enough warnings)~\cite{yu2024don}.

Generating a wallet address in an EVM-compatible chain is a three-step process:

(1) Choose a 32-byte private key ($k$),
which must be truly random to avoid a wallet breach~\cite{profanity_vulnerability}. 

(2) Apply elliptic curve (ECDSA) multiplication to produce a public key $K = k \times G$ where $G$ is the fixed generator point on the elliptic curve. 

(3) Hash (Keccak 256) the public key $K$, take the last 20 bytes, and add the prefix $0x$ to obtain the address. 

Elliptic curve (ECDSA) multiplication is computationally hard to invert, 
meaning that it is impractical to derive a private key from a public key.
To generate a public key with a certain format (e.g., a vanity
address), the only viable strategy is to perform a brute-force
computation over many private keys. The complexity of the
brute force computation depends on the number of characters that have to
be fixed in the desired public key address. For instance, it is roughly
$16^3$ times harder to generate an (hexadecimal) address with a 3-character desired prefix
compared to a random address.
Attackers adopt this brute-force technique to generate lookalike addresses. 



\subsection{Attack characteristics}
\label{subsec:attack_characteristics} 
Phishing attacks on domain names have several variants. Examples
include typosquatting~\cite{moore2010measuring}, which involves users
mistyping domain names; combosquatting~\cite{kintis2017hiding},
wherein attackers append related keywords to domain names;
and homograph attacks~\cite{gabrilovich2002homograph}, where
attackers substitute characters in a domain name with visually
similar ones (e.g., Cyrillic
characters instead of Latin~\cite{holgers2006cutting}). All these attacks exploit 
human features---our inability to distinguish between
graphically near-identical representations, as well as our 
ability to quickly make inferences from incorrect text~\cite{Rayner:PS06}. 
Domain-name-based phishing attacks may 
resort to indirect monetization, e.g., ad impressions, or affiliate program enrollment~\cite{holgers2006cutting,
moore2010measuring, agten2015seven}.

Blockchain address poisoning is markedly different from 
domain-name-based phishing attacks in several respects. 

First, domain-name based attacks often proceed in multiple stages to
steal funds: the victim first clicks on a malicious URL, is directed to
a phishing website and is prompted to input sensitive information that
can eventually be monetized~\cite{muzammil2024typosquatting}. On the
other hand, blockchain address poisoning only requires a \emph{single}
mistake from the victim---erroneously selecting an attacker address
instead of a legitimate one. Furthermore, fund recovery is far more
challenging due to the irreversibility of cryptocurrency transactions.


Second, blockchain address poisoning attempts
are permanently recorded in a public ledger, allowing for more 
accurate estimates of the losses~\cite{li2023double}. Estimating
the impact of traditional phishing attacks is challenging
due to data confidentiality~\cite{oest2020sunrise} and 
phishing website ephemerality~\cite{vasek2018analyzing,
van2019cognitive,bijmans2021catching, li2023double}.

As a third difference, address poisoning targets hexadecimal
strings, while domain-name based attacks typically focus on
character strings. Previous typosquatting studies introduced
various metrics for quantifying domain name similarity, such as
Damerau-Levenshtein (DL) distance~\cite{damerau1964technique},
fat finger (FF) distance~\cite{moore2010measuring}, and visual
distance~\cite{szurdi2017email}. However, blockchain address
poisoning involves copy-and-paste or multiple-choice selection rather than manual typing of
characters. Additionally, long hexadecimal addresses
are often abbreviated to their prefix and suffix, especially on constrained interfaces such as mobile. Thus, DL and FF distances are 
less relevant here. We instead use the number of matching hexadecimal digits
at the beginning and at the end.
 

Finally, address poisoning has differing costs to the attacker. They
only need to deploy attack smart contracts (which can be even recycled
from others for free), and one poisoning transfer costs only about a
dollar in Ethereum and a cent in BSC. Generating lookalike addresses however
has a variable cost that can be adjusted by the attacker, ranging anywhere from a laptop CPU to data centers of GPUs.
Domain-based phishing attacks, on the other hand, require registering/hijacking domains, crafting phishing
websites, and/or buying tool kits~\cite{bijmans2021catching}.

\section{The attack}
\label{sec:threat_model}
This section describes address poisoning in detail. We first
introduce our threat model and attacker capabilities, discuss
various ways to carry out address poisoning, and examine how attackers
execute them in practice.

\subsection{Threat model}
\label{sec:threat_model:description}
We assume the existence of an attacker that passively monitors
all public blockchains simultaneously in real-time.
When available the attacker may also observe a state that has yet to be
finalized in a chain, such as transactions in the mempool.
The attacker can also actively engage with any chain
by launching contracts and sending transactions using many accounts,
bounded only by the chain and economic constraints.
Finally, the attacker can generate lookalike addresses
at a cost consistent with (publicly known) state-of-the-art hardware.

We do not assume that the attacker has any out-of-band information
about or agency towards a victim, other than what can be obtained
through public blockchains. Therefore attackers cannot influence the
wallet software a victim is using, or make predictions about a victim's
transfers beyond what can be inferred through public signals (i.e.,
they cannot tailor the attack to each specific victim). We also assume
that the security models for the individual chains hold, so attackers
cannot, for example, perform denial of service on a chain or drop/modify
a victim's transactions.

The attacker's ultimate goal is to have the victim send assets to
(one of) the attacker's lookalike address(es) instead of the victim's
intended address. To do so, the attacker first monitors blockchain
transactions and identifies asset (ETH or ERC-20 tokens) transfer events.
When such transfers are observed, the attacker identifies an intended recipient wallet, $R$, and generates a
lookalike address $L$. 
We will later numerically evaluate the computational 
requirements to get lookalike addresses convincing enough for the
attack to be viable.

The attacker then submits to the chain phishing
transactions that interact with the victim $V$ 
and contain the lookalike address $L$, typically within a short period of time. 
When the victim later attempts to send
assets to $R$, they could get confused and mistakenly send funds to $L$ instead. 
This confusion typically arises through some combination of UI/UX limitations, and 
the victim failing to carefully check the recipient.
A common vector is for a victim to transfer funds to an address that is pre-populated by their wallet
software in a list of addresses with which they had recent interactions. 
We focus on ERC-20 (or BEP-2 in BSC) tokens for the underlying asset as attacks
with these tokens are most prevalent.

\subsection{Transfer types}
\label{subsec:family_of_attacks}
The attacker poisons the victim through three types of token transfers: \textit{tiny transfers}, \textit{zero-value transfers},
and \textit{counterfeit token transfers}. Figure~\ref{fig:attack_sketch}
depicts these variations.

In a tiny transfer (Fig.~\ref{subfig:tiny_transfer}), the attacker always sends
from $L$ to $V$ a small amount of the same token that $V$ had previously sent
to $R$. The idea is that $L$ now shows up in $V$'s recent transaction
history. Later, $V$ could mistakenly assign $L$
as the recipient due to its similarity to $R$.
This attack is similar to blockchain dusting~\cite{wang2018anti},
in which the attacker sends many small transactions to a victim for
deanonymization.

In a \textit{zero-value transfer} (Fig.~\ref{subfig:zero_transfer}), 
the attacker creates a transfer from $V$ to $L$ for the same token 
that $V$ sent to $R$, but with a value of zero.
Typically, token transfers can only be sender-initiated.
However, the current token standard features a public function called \textit{transferFrom}, and popular token implementations allow the function callers to specify \emph{an arbitrary sender and an arbitrary receiver} if the transferred value is zero.
Therefore, the attacker can record a transfer event from $V$ to $L$ of the same ERC-20 token with zero value.
Zero-value transfers are potentially even
more dangerous than tiny transfers, 
as the victim appears to have previously \emph{sent} money to the lookalike address.

In a \textit{counterfeit token transfer} (Fig.~\ref{subfig:fake_transfer}), 
the attacker sends from $V$ to $L$ the same amount $V$ sent to $R$, but with a counterfeit token created by the attacker (``USDT\textbf{T}'' here).
Since the attacker controls the implementation of the counterfeit token, they can have the ability to transfer it on behalf of anyone.
There is no restriction on names, symbols, or functionality for ERC-20 tokens~\cite{gao2020tracking, xia2021trade}, which makes it easy to create similar-looking tokens for malicious purposes.
Because most popular wallet software maintains a curated list of popular (chain, contract address, symbol) tuples, attackers often avoid names (or symbols) identical to mainstream tokens to avoid triggering a warning. 
In the case of new or recently launched tokens, they could even proactively register the token name first~\cite{ye2024interface}.

Finally, these types of transactions are not mutually
exclusive. Indeed, the attacker can combine these strategies, flood
the victim with transactions, and wait for a mistake.

\subsection{Attack implementation}
\label{sec:threat_model:execution}

\begin{figure}[ht]
    \centering
    \includegraphics[width=\linewidth]{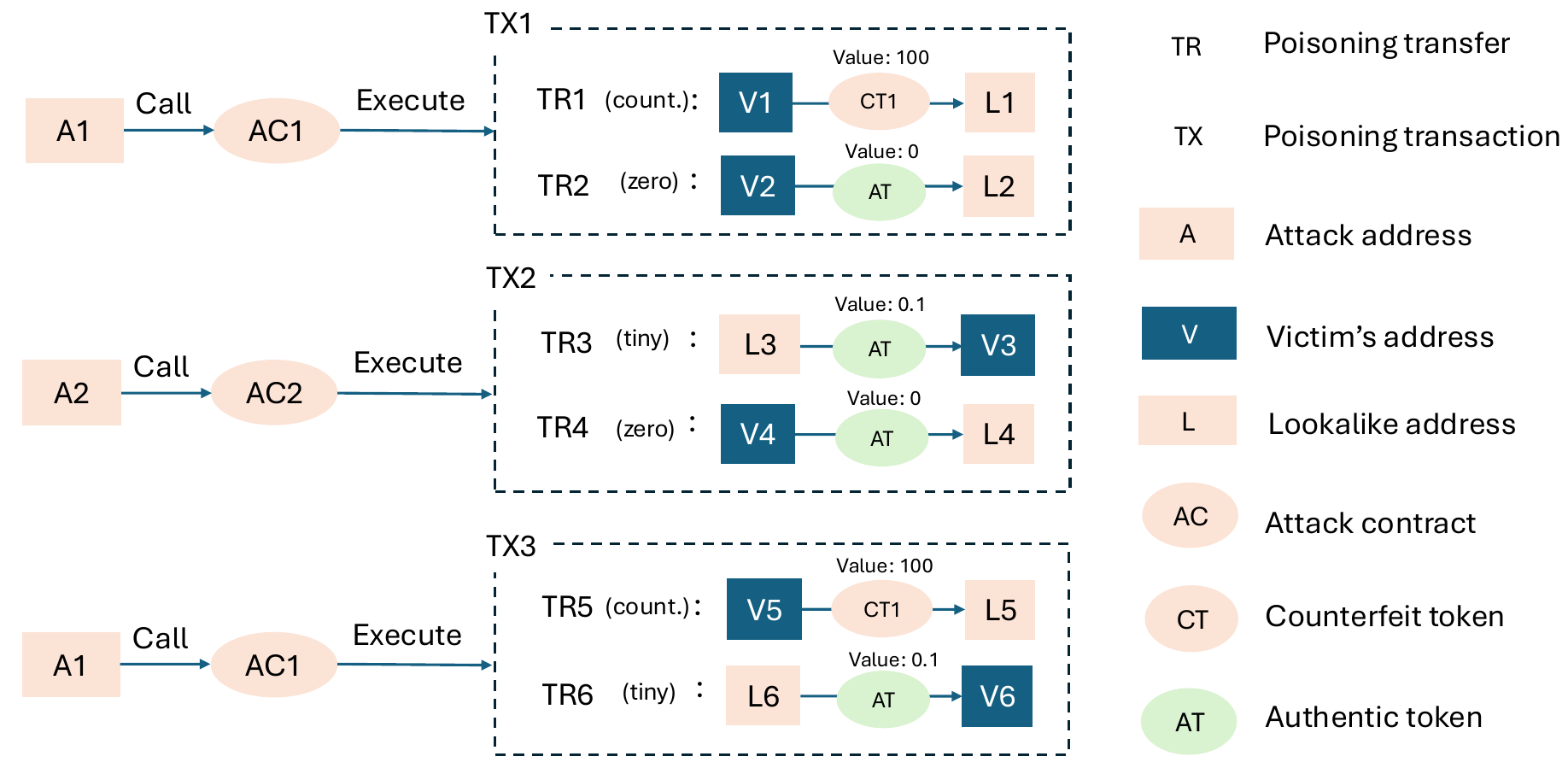}
    \caption{Attack mechanism}
    \label{fig:attack_mechanism}
\end{figure}

Figure~\ref{fig:attack_mechanism} shows one method of implementing blockchain address poisoning in practice. 
Wallets 
are represented by rectangles and contracts (or ERC-20 tokens created by contracts) are represented by ovals.
We color attackers' instances in pink. 
First, an attack account ($A$) deploys a tailor-made attack contract ($AC$) and calls a function from it to execute multiple poisoning transfers at once.
The attacker can also directly interact with the token contract to initiate a single poisoning transfer (without $AC$), which we observe in some cases despite its higher gas cost. 
The transaction $TX$ bundles together several potentially unrelated phishing transfers, in this case, the transfer of a counterfeit token from $V1$ to $L1$ and a zero-value transfer from $V2$ to $L2$. 
Notice also that attacker $A1$ recycles their contract $AC1$ and $CT1$ to perform $TX3$, and the infrastructure that $A1$ has embedded on a chain is typically distinct from that of other attackers, such as $A2$'s use of $AC2$. 
It is also common for multiple attackers to target the same victim (e.g., $V1 = V4$ ), as we will later describe.
We have observed cases where attackers launch more than 100 poisoning transfers in a single transaction.

\section{Detecting attacks}
\label{sec:method}
We next explain how we implement our detection system and how we cluster individual attack instances into attack groups.

\subsection{Detection algorithm}
\label{subsec:detection_algo}

\begin{figure}
    \centering
    \includegraphics[width=1\linewidth]{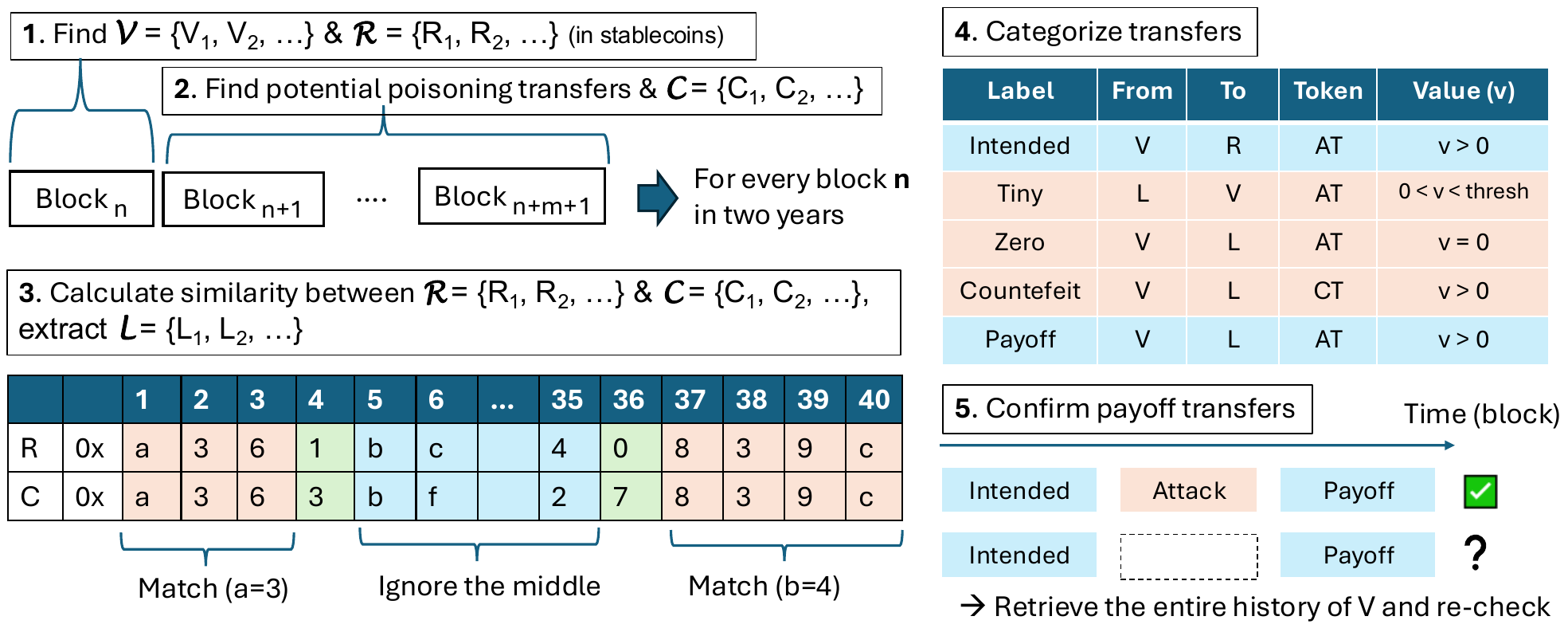}
    \caption{Detection algorithm sketch}
    \label{fig:detection_algo}
\end{figure}

Ye et al.~\cite{ye2024interface} identify zero-value transfers by enumerating all token transfers with zero value.
While computationally efficient, this method does not consider tiny and counterfeit token transfers and possibly mislabel benign zero-value transfer attacks, e.g., those used for testing a destination address.
Guan and Li~\cite{guan2024characterizing} capture all three poisoning transfers but follow an approach slightly different from ours. 
While both filter transfers when a victim interacts with two improbably
similar, our approach appears more robust to attacker manipulation (refer to \S\ref{subsec:detection_eval} and Appendix~\ref{sec:appendix:comparison}).
We deploy our own Ethereum and BSC full nodes to scan the data efficiently instead of
relying on third-party RPC endpoints. Figure~\ref{fig:detection_algo}
sketches our detection system, which consists of five steps.

\noindent\textbf{Step 1: Identifying potential victims. } 
For block~$n$, we collect the list of addresses (senders)
$\mathcal{V}=\{V_1, V_2, \ldots \}$ who have sent any amount of
one (or more) of the major stablecoins\footnote{Tether (USDT), USD
Coin (USDC), and DAI for Ethereum and Binance-Peg USD Coin (USDC),
Binance-Peg BSC-USD (BSC-USD), Binance-Peg BUSD (BUSD), and any USDC
for BSC.} and the addresses $\mathcal{R} = \{R_1, R_2,
\ldots \}$ that have received these tokens. $\mathcal{V}$ is the set of
{(potential) \textit{victim addresses}, and $\mathcal{R}$ is the set of
\textit{intended addresses}.

\noindent\textbf{Step 2: Identifying possible poisoning transfers. } 
We next collect potential \emph{poisoning transfers}, i.e., the
attempt(s) to tamper with the victim's address lists. These poisoning
transfers often happen soon after the attacker has identified a
potential victim through the original token transfer (which we collected
in Step 1). So, we focus on the 20 minutes following the
original transfer, which corresponds to blocks $n+1$ to $n+m+1$ where $m
= 100$ and $400$ for Ethereum and BSC, respectively. We validate this
choice of window size in \S\ref{subsec:detection_eval}.

We look at all transfers\footnote{We capture all transfers that emit
a ``transfer'' signature, including any contract actions.} in
stablecoins -- i.e., we search all transfers sent to (or from) addresses
in $\mathcal{V}$ for potential tiny (or zero value) transfers. We
also retrieve all ERC-20 token transfers for which an address in
$\mathcal{V}$ is a source for potential counterfeit token transfers. We
denote the set of \textit{recently used addresses} that interact with
$\mathcal{V}$ by $\mathcal{C}=\{C_1, C_2, \ldots \}$.

We iterate this two-step process for every block $n$ over
a two-year interval.\footnote{We collect transfer events
through the \texttt{eth\_getLogs} JSON-RPC API with signature
\texttt{0xddf252ad\allowbreak 1be2c89b\allowbreak 69c2b068\allowbreak fc378daa\allowbreak 952ba7f1\allowbreak 63c4a116\allowbreak 28f55a4d\allowbreak f523b3ef}.}
Given that attackers often launch multiple poisoning transfers
within a single transaction, we further retrieve other transfer
events (if any) in the same poisoning transactions. For instance, in
Figure~\ref{fig:attack_mechanism}, if we detect $TR_1$, we can also
capture $TR_2$. 

\noindent\textbf{Step 3: Extracting lookalike addresses. } 
We then extract the \textit{lookalike addresses} $\mathcal{L}$ that the attackers create to impersonate the intended addresses $\mathcal{R}$.
While researchers typically apply Hamming distance to hexadecimal strings~\cite{kintis2017hiding}, the middle part of a string matters significantly less in address poisoning.
As a result, we develop our similarity score algorithm focusing on the beginning and ending hexadecimal characters of the address.
For simplicity, we do not consider checksum addresses and regard upper and lower cases as equivalent. 
For every pair $(R_i, C_j)$ where $R_i \in \mathcal{R}$ is an intended
address and $C_j \in \mathcal{C}$, a recently used address, we check the length of the prefix and suffix
match which we record as a similarity score $(a, b)$.
To reduce false positives (i.e., addresses that are similar to each other by coincidence), we only consider the addresses in $\mathcal{C}$ for which $a \geq 3$ and $b\geq 4$.\footnote{Previous work on typosquatting or combosquatting~\cite{moore2010measuring, kintis2017hiding} avoids short domain names for the same reason.}
We choose this threshold based on the common visual representations of
blockchain addresses, which use abbreviations~\cite{ye2024interface}.
We denote the set of these lookalike addresses by $\mathcal{L} = \{L_1, L_2, \ldots \} \subseteq \mathcal{C}$. 
By the birthday paradox, if a victim interacts with many different accounts, there is a high chance of interacting with a similar-looking address by coincidence, leading to potential false positives.
Appendix \ref{sec:appendix:bday_thresh} discusses how we set the threshold to exclude victims who are highly susceptible to this issue while capturing nearly all attacks.

\noindent\textbf{Step 4: Pruning possible attacks. } 
For each victim $V$, the corresponding intended address
$R$, and the associated lookalike address $L$, we extract all related token
transfer events collected in step 2. In other words, we only look
at the cases where the victim $V$ interacts with two very similar
addresses $R$ and $L$, making false positives statistically unlikely
(except for those caused by typing mistakes, which we separately discuss in Appendix~\ref{sec:appendix:detect_accidental_transfers}).

To identify counterfeit tokens, we acquire a list of ERC-20 and BEP-20 tokens from two sources: 1) Etherscan and BSCscan\footnote{Tokens with at least a ``Neutral'' or ``OK'' reputation from Etherscan: \url{https://etherscan.io/tokens} and BSCscan: \url{https://bscscan.com/tokens} as of August 13, 2024.} 
and 2) Coinranking.\footnote{\url{https://coinranking.com/coins/erc-20} as of August 31, 2024.}
The union of these lists forms the set of authentic tokens $\mathcal{AT}$; all others form the set of counterfeit tokens $\mathcal{CT}$.

We then categorize the transfers we have collected into three major
categories (intended, poisoning, payoff), distinguishing poisoning
transfers between tiny, zero-value, and counterfeit token transfers per
their description in Figure~\ref{fig:attack_sketch}.
Let $v$ be the
monetary value of the transfer. Formally: 

\begin{itemize}
  \item A transfer is an intended transfer if a potential victim $V$ sends an authentic token $AT \in \mathcal{AT}$  to an intended recipient $R$, and $v>0$.
    \item A transfer is a poisoning transfer: 
      \begin{enumerate}
        \item If a lookalike address $L$ sends an authentic token $AT$ to a victim $V$ with $0<v<v_0$ (tiny transfer attack). 
          We set the threshold $v_0$ to 10~USD to exclude unreasonably high transfers from the attacker.\footnote{While 98.79\% of tiny transfers ($<10$ USD) are less than 3 USD, we manually confirmed some rare tiny transfers were almost 10~USD.} 
        \item Or, if a victim $V$ sends an authentic token $AT$ to a lookalike address $L$ with $v=0$ (zero-value transfer attack).
        \item Or, if a victim $V$ sends a counterfeit token $CT \in \mathcal{CT}$ to a receiver $L$ (counterfeit token transfer attack). Recall that because the token was created by the attacker, they could implement any semantics they desire, notably having anybody send tokens to anybody else, without any authentication.
      \end{enumerate}
    \item A transfer is a \emph{payoff transfer} if a victim $V$ sends an authentic token $AT$ ($v>0$) to a lookalike address $L$, which ultimately monetizes the attack for the adversary.
\end{itemize}

The top-right table in Figure~\ref{fig:detection_algo} summarizes our labeling scheme. Pink cells are attackers.
Note that we convert the transferred value into USD on the transfer date. 
The price data is from the CoinGecko API based on a contract address. 

\noindent\textbf{Step 5: Verifying payoffs.}
Among all possible payoff transfers found in Step 4, we 
consider a transfer to be \emph{confirmed} if 
$V$ has received at least one poisoning transfer from $L$ between the intended transfer and the (candidate) payoff transfer. 

Others are \emph{unconfirmed} payoff transfers\textemdash the victim sends $AT$ to $L$ without having received a poisoning transfer from $L$. This can happen for three reasons. 

First, the victim could have legitimately interacted with two similar
addresses by sheer coincidence. This is statistically unlikely,
specifically because we already filter out addresses particularly
vulnerable to the ``birthday paradox'' (refer to Appendix~\ref{sec:appendix:bday_thresh} for our detailed analysis).

Second, the attack may have taken place over a time interval exceeding
our 20-minute monitoring window. To address this, for each unconfirmed
payoff transfer, we fetch the victim's entire token transfer history
(from Etherscan and BSCscan) between the intended transfer and the
unconfirmed payoff transfer. If we find any poisoning transfers we had
failed to capture, we confirm the
payoff transfer.

Third, the victim could have accidentally sent tokens to an incorrect
destination address, e.g., due to a typing mistake. If 1) the destination 
address never sent any tokens, i.e., the attacker does not redeem the payoff transfer, and 2) $d$ is more than 20, i.e., computationally infeasible to generate such a $L$, we assume that the destination address was actually a typo. 
We indeed find 681 accidental transfers from
501 users, totaling 5.5 million USD in losses (almost all in Ethereum, with only 57,000 USD in BSC): Appendix~\ref{sec:appendix:detect_accidental_transfers} for more detailed
method, results, and analysis.

\subsection{Clustering attackers}
\label{subsec:clustering_algo}
Following traditional cybercrime literature~\cite{szurdi2014long}, we want to
link attack instances to extrapolate attack groups and analyze strategies, profitability, and infrastructure. 
Each address poisoning attack consists of 1) a poisoning transfer
($TR$), 2) a poisoning transaction ($TX$) that initiates $TR$,
3) a lookalike wallet address ($L$), 4) a counterfeit token
($CT$), 5) an attack contract ($AC$) that initiates $TX$, and
6) an attack wallet address ($A$) that interacts with $AC$ (see
Figure~\ref{fig:attack_mechanism}).

We associate each poisoning transfer $TR$ with the corresponding attack
``transfer set'' $\{TX, L, CT, AC, A\}$.
Similar to the ``guilt-by-association'' technique~\cite{xia2021trade}, we
merge two attack transfer sets if they share \emph{at least} one of $TX$, $L$,
or $A$. In other words, we assume that 1) two poisoning transfers in
the same transaction belong to the same attacker, 2) an attacker is
economically rational and specifies addresses $L$ they control (to
successfully receive payoffs), and 3) an attacker does not share the
private key of the attacker's wallet 
$A$ with other attackers. We do not use information on
$AC$ or $CT$ to avoid spurious cluster merges because anyone
can call an attack contract or use a counterfeit token, regardless of 
source code availability. 
For instance, in Figure~\ref{fig:attack_mechanism}, we 1) merge ($TR_1$,
$TR_2$) because both transfers are in the same transaction, 2) merge
($TR_1$, $TR_3$) if $L_1 = L_3$, and 3) merge ($TR_3$, $TR_4$, $TR_5$,
$TR_6$) if $A_1=A_2$.
We keep merging attack transfer sets until there is no overlap of $TX$, $L$, and $A$ between attack groups.

Having applied this process to our entire dataset, we notice that two
seemingly distinct large attack groups---active at different times or
using different attack strategies---can be merged due to a few
overlapping attack addresses $A$.
Manual investigation revealed that those accounts appear to be bots
copying poisoning transactions from two different attack groups with
little or no modification, leading to a seemingly erroneous merge.
Guan and Li~\cite{guan2024characterizing} perform similar clustering techniques, but did not consider the existence of bots and appeared to form one large cluster.   

To prevent such issues, we explored several methods to exclude
transactions from bots. While most of the approaches either still
produce erroneous merges or remove too many transactions and
underestimate the group size, analyzing account history effectively
addresses both issues. For each attack address $A$, we additionally
fetch their account history from Etherscan and calculate an \emph{attack
ratio}, defined as the proportion of transactions used for poisoning
transfers. We then set a threshold to exclude addresses that have a low
attack ratio. We document the details of those bots' behavior and the
bot exclusion process in Appendix~\ref{sec:appendix:copying_bots}.

\section{Results and evaluation}
\label{sec:results_eval}
We apply our detection algorithm over the period Jul. 1st, 2022 to Jun.
30th, 2024 for both Ethereum (blocks~15,053,226 to 20,207,948)
and BSC (blocks~19,170,674 to 40,077,592). 
The chain scanning
process (written in Go) takes over two weeks on Ethereum and over a month on
BSC (in Ubuntu 20.04.5 LTS server; AMD EPYC 9124
16-Core Processor; 128GB RAM), highlighting the need for our optimizations. 

\subsection{Detection results}
\label{subsec:detection_results}
We present the summary statistics for poisoning transfers in Table~\ref{tab:attack_stats} 
and confirmed payoff transfers in Table~\ref{tab:successful_attack_stats}.

\begin{table*}[]
  \centering
  \caption{
    \label{tab:attack_stats}
    Attack summary statistics}
  \begin{adjustbox}{width=0.95\textwidth,center}
    \begin{tabular}{@{}lcccccccccc@{}}
      \toprule
            & \multirow{2}{*}{\textbf{Blocks}}     & \multirow{2}{*}{\textbf{Transactions}} & \textbf{All poisoning} & \textbf{Tiny} & \textbf{Zero-value} & \textbf{Counterfeit-token} & \textbf{Victim}     & \textbf{Lookalike} & \textbf{Attack} & \textbf{Counterfeit} \\
            & & & \textbf{transfers} & \textbf{transfers} & \textbf{transfers} & \textbf{transfers} & \textbf{addresses} & \textbf{addresses} & \textbf{contracts} & \textbf{tokens} \\\midrule
            Ethereum   & 5,154,722  & 1,691,529    & 17,365,954                 & 308,881        & 7,185,298            & 9,871,775                   & 1,330,948        & 6,492,215           & 3,480            & 6,280              \\
            BSC   & 20,906,918 & 16,505,215   & 252,703,515                & 3,651,015      & 140,556,905          & 108,495,595                 & 16,107,774       & 43,644,433          & 406              & 710                \\
            Total & 26,061,640 & 18,196,744   & 270,069,469                & 3,959,896      & 147,742,203          & 118,367,370                 & 17,438,722       & 50,136,648          & 3,892            & 6,990              \\ \bottomrule
            
    \end{tabular}
  \end{adjustbox}
\end{table*}

\begin{table}[]
  \centering
  \caption{
    \label{tab:successful_attack_stats}
    Successful phishing attacks}
  \begin{adjustbox}{width=\columnwidth,center}
    \begin{tabular}{@{}ccccccccc@{}}
    \toprule  
      & \textbf{Total Loss} & \textbf{Min}   & \textbf{Median} & \textbf{Avg.}     & \textbf{Max}        & \textbf{Std. dev}     & \multirow{2}{*}{\textbf{Count}} & \multirow{2}{*}{\textbf{Victims}} \\
      & (USD)  & (USD) & (USD) & (USD) & (USD) & (USD) \\ \midrule
      Ethereum & 79,344,412 & 0.001 & 2,160  & 45,653 & 20,000,000 & 514,142 & 1,738 & 1,502   \\
      BSC & 4,490,804  & 0.000 & 85     & 1,164  & 279,489    & 7681   & 4,895 & 4,004   \\ 
      \bottomrule
    \end{tabular}
  \end{adjustbox}
\end{table}

On Ethereum, we find 17.3M poisoning transfers: 300,000 tiny transfers, 7.2M zero-value transfers, and 9.9M counterfeit-token transfers over 1.7M transactions.
Attackers target
1.3M victim addresses from 6.5M lookalike addresses. 
On Feb. 18th, 2023, we capture 362,934 poisoning transfers 
($\approx$50 transfers per block).
Those attack transactions consume 6.6\% of total gas used that day, producing wasteful state on the chain and increasing network operating cost\footnote{We assume the average block gas limit of 15 million, and aggregate gas usage of all poisoning transactions to calculate the percentage.}. 

Attackers on Ethereum receive 1,738 transfers from 1,502 victims, totaling nearly 80M USD (avg. 45,853 USD, median 2,169 USD, with high variance), with a success rate is 0.01\%\footnote{0.13\% spam click-through rate in Twitter~\cite{grier2010spam}}.
Some victims sent assets to the attacker's address(es) multiple times.
For instance, one victim sent 1.999M and 2M USDC within 10 blocks without noticing they were being phished.

On BSC, the number of attacks is significantly larger. 
We identify over 252M poisoning transfers: 3.6M tiny transfers,
141M zero-value transfers, and 108M counterfeit-token transfers in
17M transactions.
Attackers target 16M victim addresses from 44M lookalike addresses. 
On Jun. 5th, 2024, we find over 3M poisoning transfers (i.e., $\approx$105 transfers per block).
The result suggests a higher attack prevalence in chains with lower transaction fees, leading to a significant clutter in UIs and degradation in user experience.
Attackers on BSC receive 4,895 transfers from 4,004 victims, totaling 4.5M USD (avg. 1,164 USD, median 85 USD).

As discussed in \S\ref{sec:method}, we only focus on a few
stablecoins for data collection efficiency, so our estimate is a
conservative lower bound. For instance, in a well-known address
poisoning attack, a victim mistakenly sent 1,155 WBTC ($\approx$68M USD) to a lookalike address in Ethereum \cite{chainalysis2024anatomy, WBTCPoison}. We
capture the attack attempt (which belongs to the 13th largest group in
our clustering results in Section~\ref{subsec:clustering_results}) but
we do not capture the final transfer (i.e., WBTC which is not a stablecoin), so our loss estimates
do not include this attack.

\subsection{Detection evaluation}
\label{subsec:detection_eval}

\noindent\textbf{Evaluation with external dataset.}
For our detector evaluation, we obtained from Guan and
Li~\cite{guan2024characterizing} their evaluation dataset and detection
results.\footnote{The authors kindly made the evaluation open source
recently: \url{https://github.com/DS2L/Poison-Hunter}.}
The data consists of 5,890 lookalike addresses and 1,154
benign (non-lookalike) addresses. The authors include lookalike
addresses only when two sources (Forta and Etherscan) report them
and label DeFi entities as benign based on the Etherscan Label
Cloud.\footnote{\url{https://etherscan.io/labelcloud}}
We only use our results from Nov. 2022 to Feb. 2024 to align with
their experiment. We discovered and manually confirmed that
45 addresses are mislabeled as lookalike---there are no corresponding poisoning
transfers. We have contacted Guan and Li, who agreed to remove these
mislabeled entries from their dataset. 
After removing mislabeled entries,
our detector achieves 97.7\% accuracy, 100\% precision, 97.2\% recall,
and a 98.58\% F1-score. In absolute terms, we identify 5,681 lookalike
addresses with 164 false negatives and 0 false positives.

We investigate those 164 false negatives using Etherscan.
A full discussion of the whole analysis can be found in
Appendix~\ref{sec:appendix:analyze_fn}. Here we summarize the high-level
results.
In 87 instances of false negatives, we fail to identify the lookalike   
address because the victim engaged in a benign transfer that resembles  
a poisoning transfer. For instance, the victim may have sent a          
legitimate zero-value transfer to the intended address for testing.  
Our detector cannot distinguish between that case and that of an
attacker creating a zero-value transfer appearing to originate from
the victim, and sent to a lookalike address. In such a case, our detector
conservatively does not classify the event as an attack, and tags the
corresponding lookalike address as legitimate. 23 false negatives are
due to attacks being outside our time-window range (i.e., attackers
target the victim after 20 minutes). 16 false negatives come from
attackers targeting other tokens (e.g., TrueUSD, BUSD) while we only
look at major stablecoin transfers. The rest (38 false negatives) can be
primarily attributed to data collection quirks. Importantly, our choice
of similarity threshold $(3, 4)$ or that our detector only considering
senders as potential victims is not the cause of any of the false
negatives.

Guan and Li's detector discovers 144 new addresses we did not catch,
while our detector captures 96 addresses that Guan and Li's did
not. Guan and Li's detector employs additional checks using attack
characteristics (e.g., checking a transaction initiator) and filters out
benign transfers that resemble  
poisoning transfers (i.e., mitigating the first case of false negatives we
identified above). However, Guan and Li's approach (slightly overfitting
to known attacks) may miss poisoning transfers when the attacker
modifies their attack strategy (e.g., using a lookalike address to
launch a poisoning transaction).

\noindent\textbf{Manual evaluation on successful cases and attack contracts. }
Statistically, the probability of getting a false positive is exceedingly low because 1) we only look at transfers where the victim has interacted with two similar addresses, and 2) we observe poisoning transfers between an intended and a payoff transfer take place. 
Nevertheless, two of the authors also manually verified transfers labeled as successful payoff transfers.
We sorted payoff transfers by decreasing the amount and checked the 30~largest transfers for both chains.
We visited each victim's page on external blockchain scanning websites, Etherscan and BSCscan, and manually confirmed 1) the intended (original) transfer, 2) the poisoning transfer, and 3) the payoff transfer and their order. 
We verified that all 30~cases are indeed payoff transfers for both Ethereum and BSC (i.e., no false positives).  
Those 30~cases account for 53.5M (67\%) of Ethereum losses and 2M (45\%) of BSC losses.

We perform the same operation for the contracts used for poisoning transfers ($AC$).
We sort $AC$ based on the number of poisoning transfers, manually look at the top 30 $AC$ for each chain on external blockchain scanning websites, and confirm at least one poisoning transfer (and its corresponding intended transfer on the victim's page).
We verify that all the contracts are indeed attack contracts, and they account for 4.9M (28\%) on Ethereum and 210.1M (83\%) on BSC.

\noindent\textbf{Checking the contract verification. }
We may mislabel counterfeit token transfer attacks if a transferred token is authentic but not in our ERC-20 token list ($\mathcal{AT}$). 
In general, public ERC-20 token creators upload the smart contract source code to Etherscan, get verified by Etherscan, and allow users to interact with (e.g., call a function of) that contract.
Attackers, on the other hand, have few incentives to upload their source code. 
By leveraging this fact, we check the existence of the smart contract source code (i.e., verification) on Etherscan for each counterfeit token $CT$.
We also examine each attack contract $AC$ for the same reason.
7 (out of 6,280) counterfeit token contracts and 8 (out of 3,480) attack contracts are verified on Etherscan. 
These involve only 14 out of 17.3M poisoning transfers. 
Such potential false positives could happen when the victim specifies a similar address (by a typing mistake or by coincidence) \textit{and} uses a very minor (but not counterfeit) token (i.e., $\notin \mathcal{AT}$) at the same time.
Some of those verified contracts are false positives (e.g., authentic tokens that are not in our dataset), but others are indeed counterfeit contracts that attackers uploaded to Etherscan. 
In other words, some attackers had uploaded the source code of the attack contracts---incidentally, this helped us illuminate how the attack is conducted (which we described in \S\ref{sec:threat_model}). 
We exclude verified attack contracts and counterfeit token contracts from our clustering to avoid potential erroneous merge. 

\noindent\textbf{Parameter selection. }
We set key assumptions about the window size (100~blocks) and the
similarity threshold (3, 4) to make the chain scanning process efficient
and reduce false positives. We evaluate those assumptions by collecting
data over 200,000 blocks starting from 16,950,603 (Apr. 1st, 2023), and
performing a detection with a window size of 200 and similarity score
of (3, 3). Table~\ref{tab:parameter_selection} compares the different
parameters and summarizes the results on the number of poisoning
transfers, lookalike addresses, and payoff transfers.

The 200-block range captures 1,363 (0.17\%) more poisoning transfers,
1,139 (0.27\%) more lookalike addresses, and the same number
of payoff transfers. Hence, doubling the window size does not
dramatically increase the number of attacks we discover. We believe
that collecting other transfers within the same poisoning transaction
(\S\ref{subsec:detection_algo}: Step 2) has significantly helped the
coverage.

When we select the threshold of (3, 3) instead of (3, 4), we capture
2,937 (0.36\%) additional poisoning transfers and 972 more lookalike
addresses (0.23\%). However, if we decrease the threshold
by 1, we are $\approx \sqrt{16}=4$ times more likely to find benign
lookalike addresses. One additional payoff transfer
identified with a threshold of (3, 3) was indeed a false positive. In
short, changing those parameters does not appear to outweigh the cost of
data collection and false positives.

\begin{table}[]
  \centering
  \caption{
    \label{tab:parameter_selection}
    Detection results with different parameters. T: tiny, Z: zero-value, C: counterfeit token transfer, L: lookalike address}
  \scalebox{0.75}{
  \begin{tabular}{@{}ccccc@{}}
  \toprule
  \textbf{Window size} & \textbf{Similarity} & \textbf{Poisoning (T, Z, C)} & \textbf{L} & \textbf{Payoff} \\ \midrule
  100 & (3, 4)     & 807,402 (250, 102,154, 704,998)  & 427,317         & 90              \\
  200 & (3, 3)     & 808,765 (252, 102,181, 706,332)  & 428,456         & 90              \\
  100  & (3, 3)     & 810,339 (265, 102,363, 707,711)  & 428,289         & 91              \\ \bottomrule
  \end{tabular}
  }
\end{table}

\subsection{Clustering results}
\label{subsec:clustering_results}
Before clustering, we attempt to remove addresses (bots) that copy transactions by setting an attack ratio threshold, as discussed in \S\ref{subsec:clustering_algo}.
We choose a threshold of 0.5, which 1) avoids erroneous merge of large clusters and 2) only removes 1.49\% of all transfers. 
We provide a detailed explanation of this threshold selection in Appendix~\ref{sec:appendix:copying_bots}. 

In total, we identify 49 groups (with more than one lookalike address), corresponding to 97.4\% of poisoning transactions, corroborating our detection results. 
Table~\ref{tab:attack_group_general} describes the number of
attack instances and assigns Group IDs sorted by the number of
lookalike addresses in each group on Ethereum, following the notation in \S\ref{sec:threat_model}. 
We only list the top eight groups
because the eighth group is significantly smaller than
the top seven (i.e., less than 100,000 lookalike addresses with a
few successful cases). Different groups exhibit different attack
strategies; not all groups use all three poisoning types. Only Group~1 uses tiny transfers. Groups 3, 4, and
7 do not use any counterfeit token transfers, and instead focus on zero-value
transfers.
Additionally, we collect contract bytecode from our Ethereum node and list
the number of distinct contracts in parentheses. In Group~2, for
example, all $CT$ and $AC$ are identical, indicating the same source code reuse (but with different addresses). 
Given that the contract bytecode changes even with a slight alteration in the source code, the high number of identical contracts corroborates the validity of our clustering
techniques (especially since our clustering does not rely on those
contracts).

\begin{table}[]
\centering
\caption{
  \label{tab:attack_group_general}
  Attack group statistics. Notations follow Fig.~\ref{fig:attack_mechanism}. The distinct number of contracts is in parentheses.}
  \begin{adjustbox}{width=\columnwidth, center}
    \begin{tabular}{@{}lrrrrrrr@{}}
      \toprule
      \textbf{Group} & 
      \multicolumn{1}{c}{$\mathbf{L}$} & 
      \multicolumn{1}{c}{$\mathbf{CT}$} & 
      \multicolumn{1}{c}{$\mathbf{A}$}  & 
      \multicolumn{1}{c}{$\mathbf{AC}$} & 
      \multicolumn{1}{c}{$\mathbf{R}$} & 
      \multicolumn{1}{c}{$\mathbf{TR}$} & 
      \multicolumn{1}{c}{$\mathbf{TX}$}      \\ \midrule
      1     & 1,390,902 & 3,309 (517) & 4,818 & 2,147 (67) & 1,091,130 & 6,485,270 & 640,600 \\
      2     & 1,221,956 & 1,491 (1)   & 382   & 259 (1)    & 970,832   & 3,329,931 & 573,500 \\
      3     & 1,199,268 & 0 (0)       & 241   & 242 (2)    & 833,140   & 2,317,361 & 127,215 \\
      4     & 473,140   & 0 (0)       & 27    & 27 (4)     & 308,079   & 595,720   & 38,675  \\
      5     & 465,389   & 392 (149)   & 241   & 227 (74)   & 375,444   & 1,003,463 & 15,222  \\
      6     & 423,107   & 81 (46)     & 104   & 13 (13)    & 280,045   & 1,303,112 & 63,856  \\
      7     & 272,557   & 0 (0)       & 27    & 39 (39)    & 272,559   & 600,876   & 16,368  \\
      8     & 99,020    & 43 (11)     & 399   & 3 (2)      & 75,742    & 210,831   & 13,589  \\ \bottomrule
    \end{tabular}
  \end{adjustbox}
\end{table}

We also investigate the number of poisoning transfers over time. Rather
than focusing on the aggregate number, we present the number of attacks
from each group. Figure~\ref{fig:attack_group_tx_week} illustrates the
number of poisoning transfers (not transactions) per week from the top
seven groups. We started to observe attacks from Dec. 2022, starting
with Group~7. Early in 2023, Groups~3 and 4 enter the market and Group~1
comes around Feb. 2023. While several top groups stop operations around
May 2024 (in Figure~\ref{fig:attack_group_tx_week}, Ethereum), the
attack is still active today; we observe a drastic increase in BSC in
June 2024 (in \S\ref{subsec:detection_results}). As we will see in
Table~\ref{tab:attack_group_profit}, we could not confirm an early mover
advantage (i.e., later comers are still profitable). The variation
in the attack strategies and the active periods across attack groups
further supports our clustering results.

All in all, we consistently see behavioral heterogeneity across
the groups we identified, such as distinct attack strategies,
identical contract reuse, and different activity periods. In
Appendix~\ref{sec:appendix:copying_bots}, we also demonstrate the
robustness of our clustering results through temporal clustering; most
attack groups stay stable/distinct over time.

\begin{figure}
    \centering
    \includegraphics[width=1\linewidth]{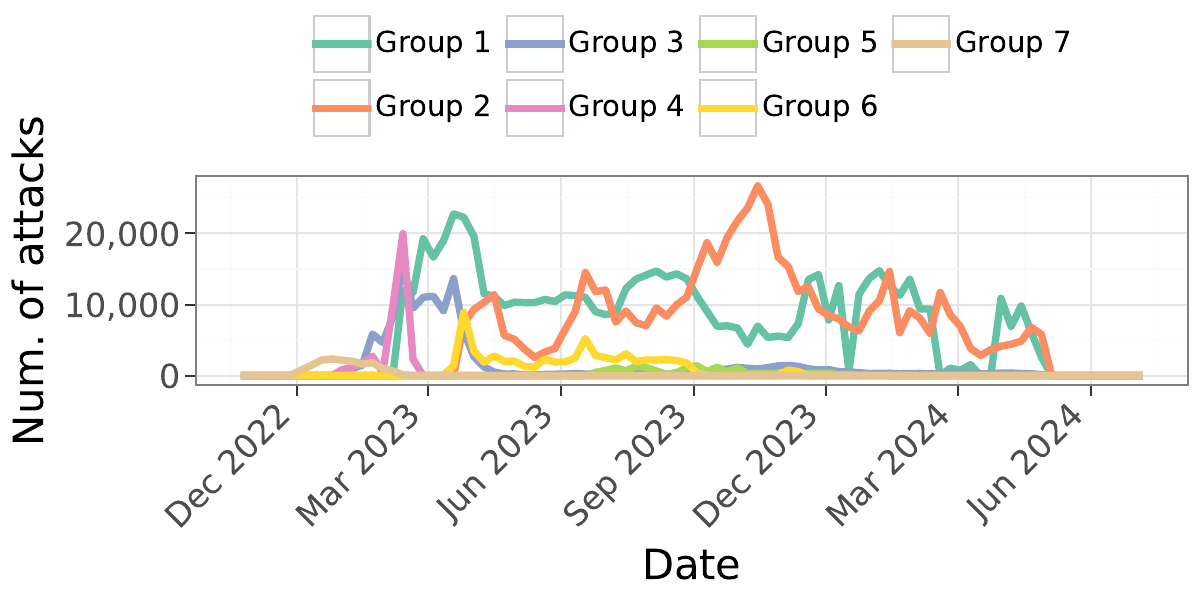}
    \caption{Number of poisoning transfers from each attack group over time (weekly basis)}
    \label{fig:attack_group_tx_week}
\end{figure}

\section{Analysis}
\label{sec:analysis}
We next analyze the detection outcome to 
identify the population(s) targeted, illuminate attacker strategies, 
investigate success conditions, and estimate attack profitability.

\subsection{Targeted population}
\label{subsec:targeted_population}
Attackers appear to select their targets in advance because the
attack typically happens right after the target's original transfer
(mostly within 20 minutes, as shown in \S\ref{subsec:detection_eval}).
We further investigate whether attackers disproportionately target
certain populations compared to the average stablecoin user.

To make the
attack more profitable, we conjecture that attackers are more likely to
target rich addresses. As different attack groups may employ different
strategies, we look at victims on a group-by-group basis on Ethereum. 
For this purpose, we maintain an Ethereum archive node that retains all historical account states.
Specifically,
we randomly sample 1,000 victims from each of the top seven attack groups. For
each victim, we retrieve the sum of their three stablecoins (USDT, USDC,
DAI) balance just before when the victim receives the first poisoning
transfer from each attack group.
We also create a baseline group by randomly sampling 1,000~USDT users
and looking up their balance (for three stablecoins) before the most recent transfer. Figure~\ref{fig:cdf_victim_stable_coin_balance} illustrates the
cumulative distribution function (CDF) of the total stablecoin balance for
the victims in each of the seven attack groups and the baseline USDT users.
All attack groups tend to target addresses that have significantly more
stablecoins than regular USDT users. Groups~6 particularly target rich
accounts given that most of the victims own more than 10,000 USD at the
time of the attack. 
Targeting rich accounts seems successful since we detect 83 payoff transfers with more than 100,000 USD.

\begin{figure}
    \centering
    \includegraphics[width=1\linewidth]{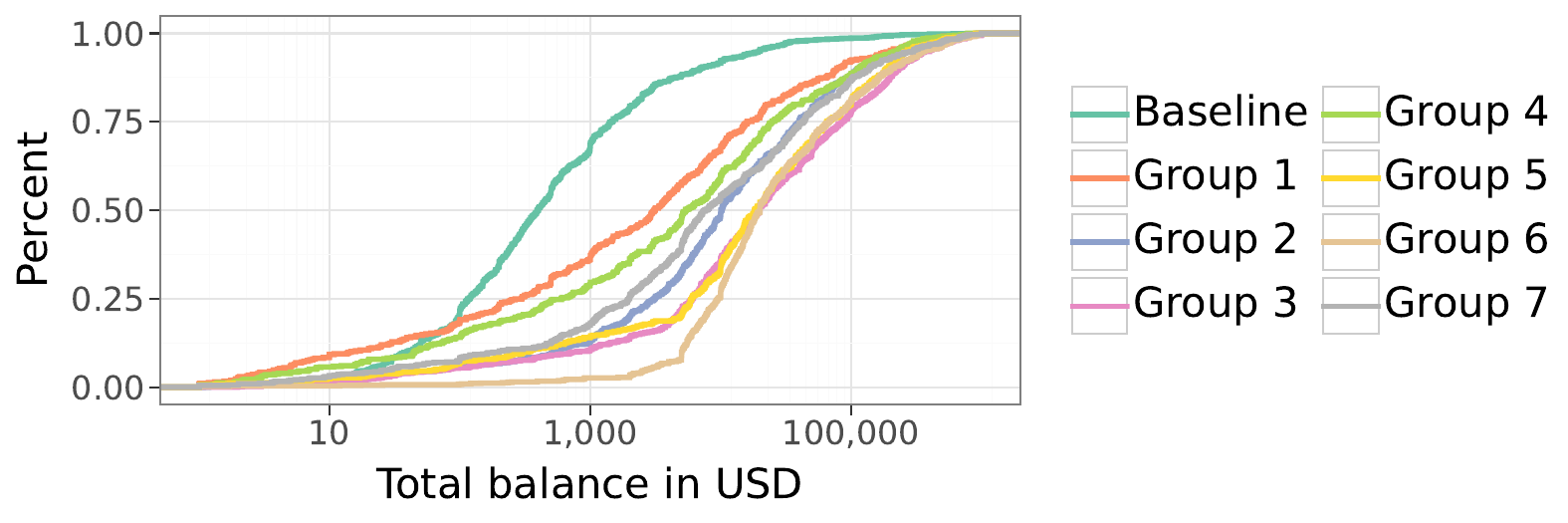}
    \caption{CDF of targeted victims' stablecoin balance for the top 7 groups (and the baseline). $x$-axis is in logscale.}
    \label{fig:cdf_victim_stable_coin_balance}
\end{figure}

Attackers also appear to target addresses that are active or have
transferred a large amount. To verify this, we sample three
sets of 100,000 accounts: 1) the most active USDT users in terms of number of transfers,
2) the users who have made the largest USDT transfers and 3) a baseline of randomly
selected USDT users. We calculate the number of attacks each user
receives. Figure~\ref{fig:cdf_num_attacks_three_population} is the CDF
of the number of attacks for three groups. Active users and large 
transfer users appear to significantly attract more attacks than other users.

Within the targeted victim population, we confirm that the larger the number (or the amount) of transfers they are involved in, the more they are targeted. Specifically, for every
victim, Spearman's rank correlation (robust to outliers) between
the number of stablecoin transfers a victim has made and the number
of attacks received is $\rho = 0.7$. Spearman's rank correlation
between the maximum amount of a transfer and the number of attacks
received is $\rho' = 0.45$.
All of those
results support that attackers appear to target users with 
1) high activity, 2) large transfers, and 3) high balance.

\begin{figure}
    \centering
    \includegraphics[width=1\linewidth]{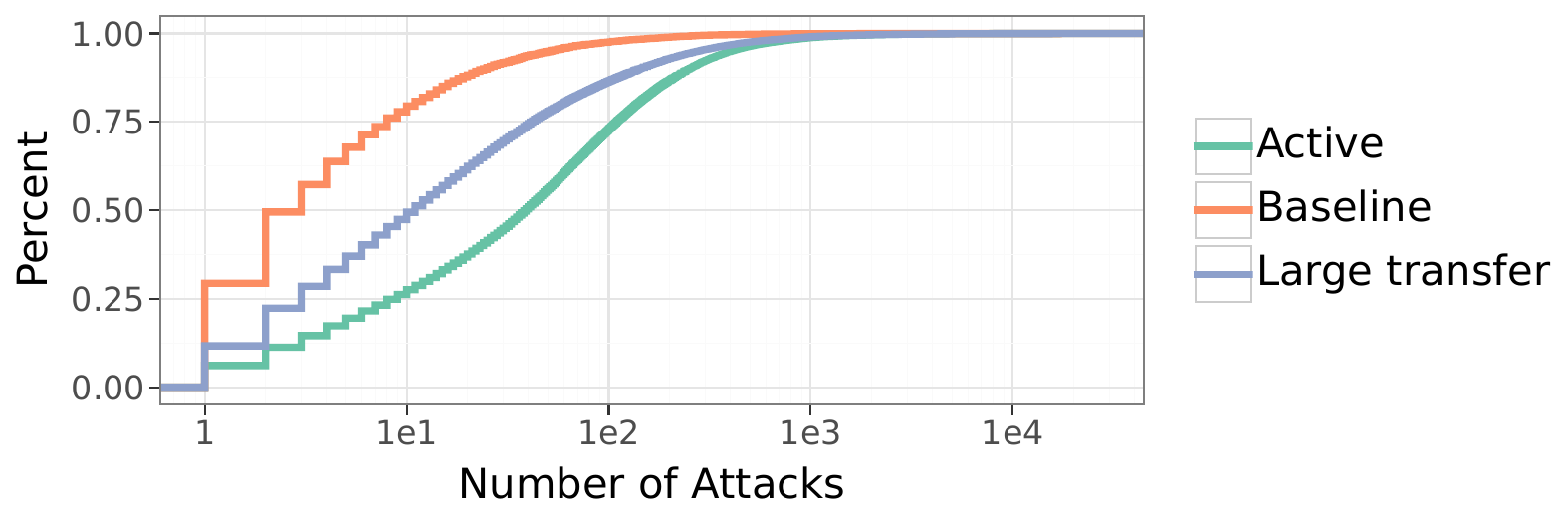}
    \caption{CDF: the number of attacks received for three groups: active, large transfers, baseline. logscale $x$-axis. The more the curve is to the right, the more the group is targeted.}
    \label{fig:cdf_num_attacks_three_population}
\end{figure}

\subsection{Attack strategies}
\label{subsec:attack_strategies}
We next infer attack strategies. 
In particular, we investigate how attackers select which (and how similar) lookalike addresses to generate and target multi-chain addresses. 

First, we discover that the attacker tends to imitate addresses ($\mathcal{R}$) from centralized entities. 
By manually looking at the list of addresses in $\mathcal{R}$ on Ethereum that the attacker imitates the most, we find that those addresses often belong to centralized exchanges (e.g., Coinbase, Binance, Bybit), probably because they are one of the most active accounts.
However, most addresses are hot wallets, with which normal users typically do not interact. 
Indeed, we do not find any successful attack involving those hot wallets.
This appears to
be a waste of time and money, as excluding such addresses would reduce
address generation and transaction costs.
We have more detailed information in Appendix~\ref{sec:appendix:intended_addr}.

Second, we observe distinctive patterns in generating lookalike addresses.
Figure~\ref{fig:fake_addr_similarity_dist} illustrates the distributions of lookalike addresses based on the number of digits matched at the beginning of the string ($x$-axis) and the end ($y$-axis) for Group~1 (left) and all other groups (right). 
The bubble size represents the number of lookalike addresses. 
The difficulty of matching addresses exponentially increases by a power of 16 for each additional hexadecimal digit (see details in \S\ref{sec:address_generation}). 
For most of the groups (the right figure), the bubble size decreases when $x$ and $y$ increase, forming a triangle shape starting from $(x, y)=(3, 4)$ as expected. 
For Group~1, we observe an increase in the number of lookalike addresses starting from $(x, y)=(7, 6)$ (i.e., the second triangle), indicating that Group~1 seems to use two different address generation methods.
One possible conjecture is that they target different services. 
Most wallet software (e.g., MetaMasks, Phantom Wallet) used to show only the first and last 3--5 characters, while blockchain scan services (e.g., Etherscan) used to display the first and last 6--7 characters.

\begin{figure}
  \begin{subfigure}{0.48\linewidth}
    \includegraphics[width=\textwidth]{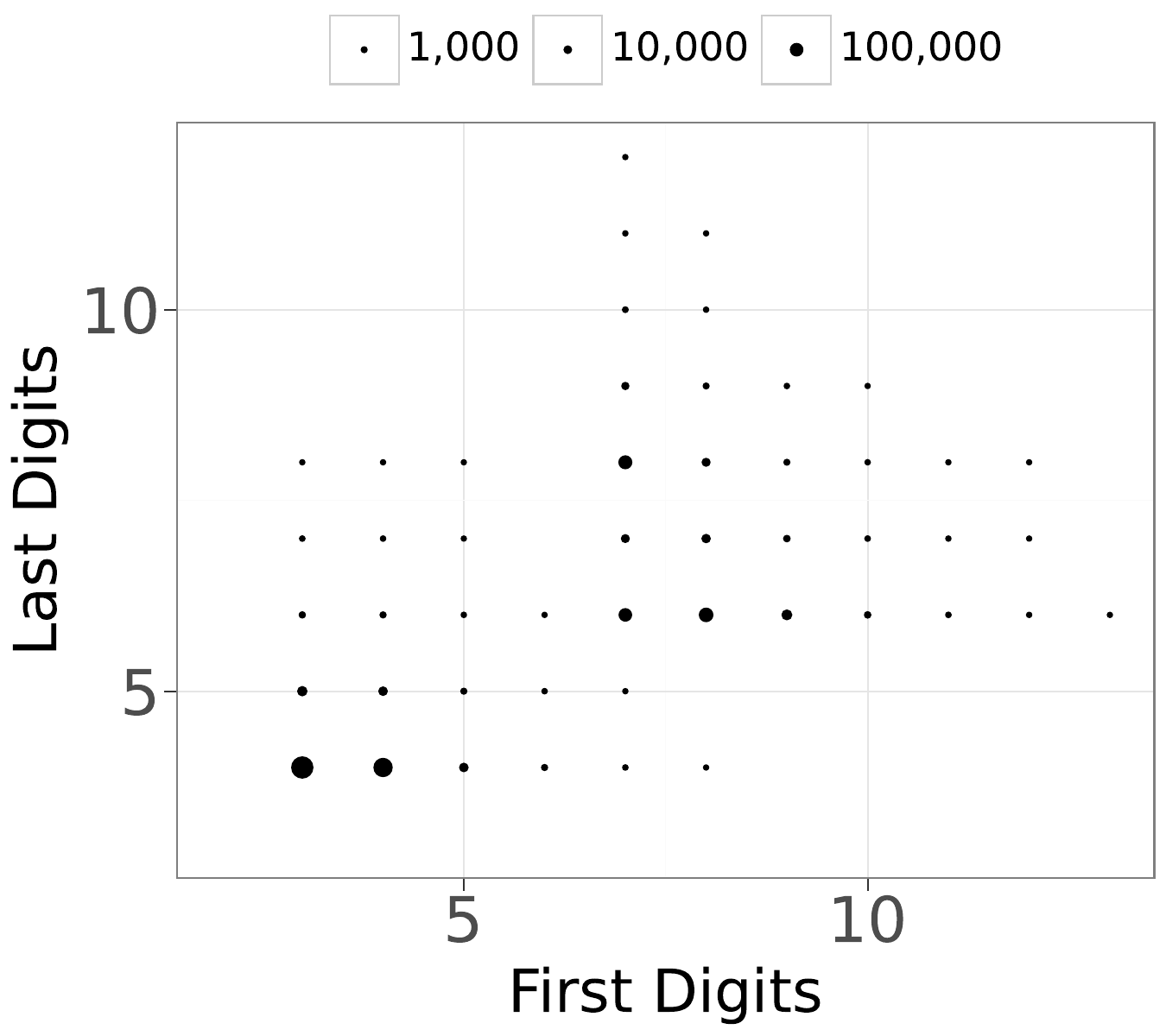}
    \caption{Group~1}
  \end{subfigure}
  \begin{subfigure}{0.48\linewidth}
   \includegraphics[width=\textwidth]{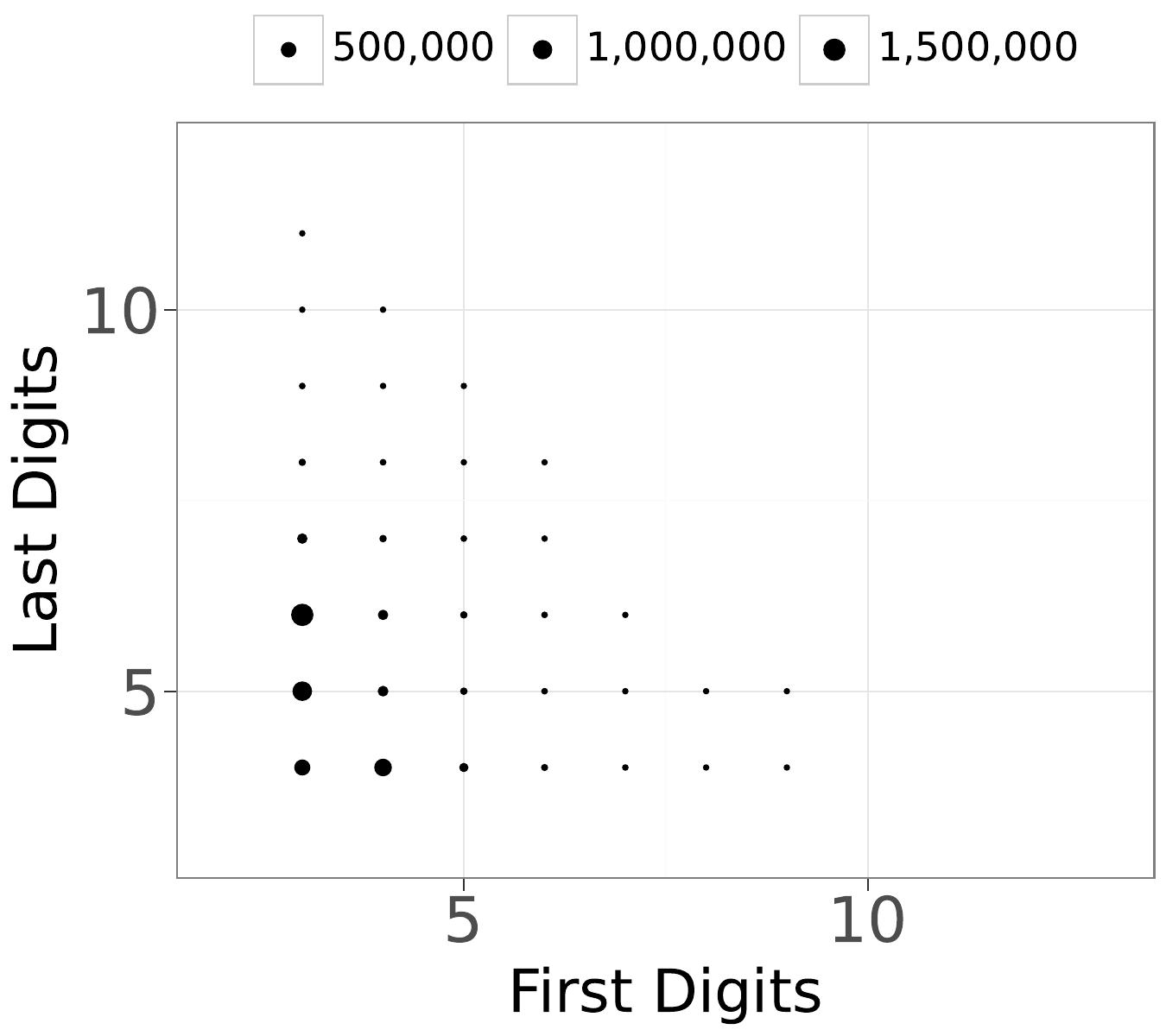}
   \caption{Other groups}
  \end{subfigure}
  \caption{
    \label{fig:fake_addr_similarity_dist} 
    Number of lookalike addresses for which the first $x$- and the last $y$-digit match the intended address.}
\end{figure}

Figure~\ref{fig:attack_group_fake_addr_dist} depicts the distributions of lookalike address similarity for the top four groups. 
While most groups generate addresses with a maximum of 14~matching digits, Group~1 manages to produce addresses with 20~matching digits, which requires higher computational power by several orders of magnitude. 
The difference in distributions also suggests that different attack groups may have employed different infrastructures, which we will investigate in \S\ref{subsec:simulating_addr_generation}. 

\begin{figure}
    \centering
    \includegraphics[width=0.87\linewidth]{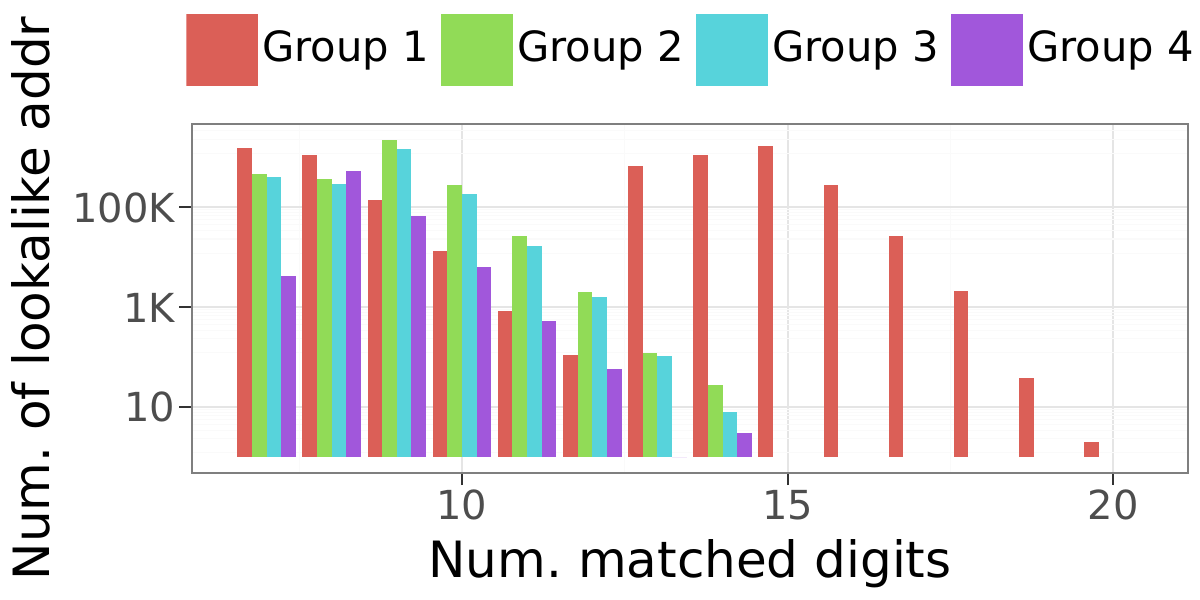}
    \caption{Distributions of the lookalike address similarity (for the top 4 attack groups): y-axis is on a log scale.}
    \label{fig:attack_group_fake_addr_dist}
\end{figure}

Finally, we investigate whether any attack group re-uses lookalike addresses across different chains.
For any EVM-compatible chains, users can use the same wallet addresses across different chains.
Attackers can reduce computation 
especially as the most active accounts frequently hold (ERC-20 and BEP-20) tokens across chains.
We observe that attackers re-use 16,903 lookalike addresses and target the same 107,542 victims across two chains.  
In particular, Group~4 re-use 21, Group~6 re-use 317, and Group~7 re-use 14,778 lookalike addresses between Ethereum and BSC, while the other groups do not.
These results imply that the attack can easily scale up to any EVM chain. 
More generally, blockchain address poisoning can affect any blockchain 
whose addresses are represented by strings or objects complex enough that a user cannot easily spot minor differences. 
For example, there have been attacks on blockchains, such as
Tron,\footnote{\url{https://support.token.im/hc/en-us/articles/12967949725593-Security-Alert-0-USDT-transfer-scam}} which do not use a
hexadecimal addressing system.

\subsection{When is the attack successful?}
\label{subsec:success_conditions}
We next look into which conditions are satisfied for an attack to be successful. 
We test our hypothesis that the more similar the lookalike address is to the intended address, the more likely the victim will make a mistake.
In Figure~\ref{fig:fake_addr_similarity_dist_success_prob}, the bubble size represents the success ratio (the number of successes to the number of lookalike addresses) depending on similarity.
We do not consider the number of attack attempts for each victim. 
While $(x,y)=(7, 10)$ and $(x,y) = (9, 8)$ have high success rate, 
there is no strong linear correlation between similarity
and success ratio in general.
This result suggests that the attack may have been more profitable to generate a larger number of slightly less convincing lookalike addresses and conduct more attacks.

\begin{figure}
    \centering
    \includegraphics[width=0.7\linewidth]{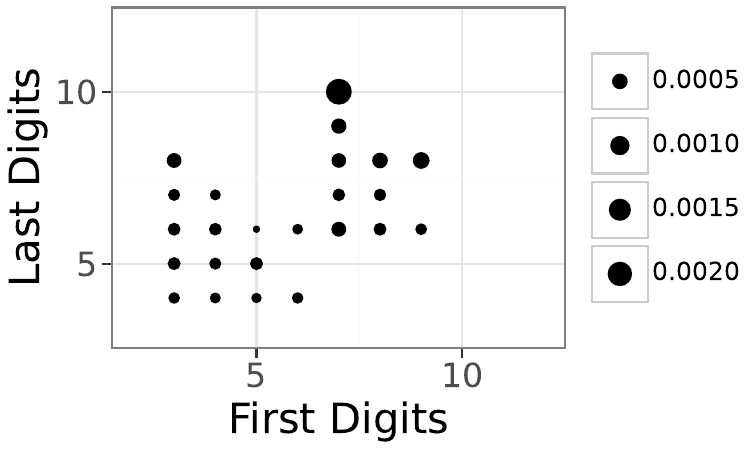}
    \caption{Success probability and address similarity. The first $x$- and the last $y$-digit match the intended address.}
    \label{fig:fake_addr_similarity_dist_success_prob}
\end{figure}

We next focus on \textit{successful cases} and examine how address similarity or attack timing relates to success. 
Figure~\ref{fig:hist_competitions} (left) shows the number of lookalike addresses targeting each victim. 
In 79\% of the successful attacks on Ethereum, multiple lookalike addresses target one victim, indicating competition among attackers.
This competition allows us to investigate how each victim selects one particular lookalike address, a winner $L_{\text{win}}$, over a pool of other lookalike addresses, while partially controlling for individual user attributes.
We focus on two aspects: how similar the winner $L_{\text{win}}$ is to $R$; and how early the winning address appears, 
compared to other lookalike addresses. 
For each successful attack, we extract the list of corresponding lookalike addresses $\mathcal{L}$, then check how $L_{\text{win}}$---the recipient of the payoff transfer---ranks in these two metrics. 
A smaller rank implies a higher similarity and an earlier attack, respectively.
Figure~\ref{fig:hist_competitions} shows that, while the number of      
competitors (left) decreases somewhat linearly, a disproportionately   
large proportion of the most similar addresses (center), and/or          
early-comers (right) tend to win the competition. 
In other words, in times of stronger  
competition, being the first and/or having a lookalike address most     
similar to the intended recipient seems to yield measurable benefits.   

While those results provide insights from chain activities, the success also relies on what wallets or chain scanning services victims use.
Appendix~\ref{sec:appendix:ui_design} compares how poisoning attacks appear
in those services' UI and qualitatively discusses when and where victims
are most likely to fall for the attack. 

\begin{figure}
  \begin{subfigure}{0.32\linewidth}
    \includegraphics[width=\textwidth]{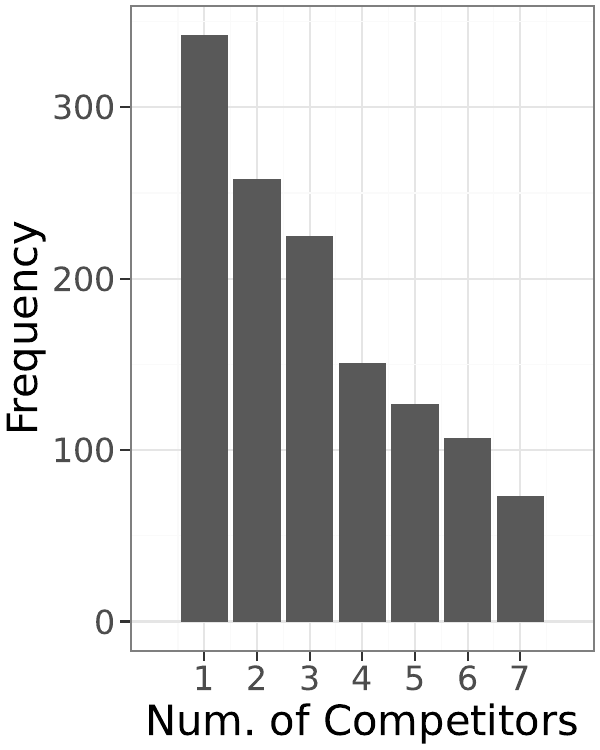}
  \end{subfigure}
  \begin{subfigure}{0.32\linewidth}
   \includegraphics[width=\textwidth]{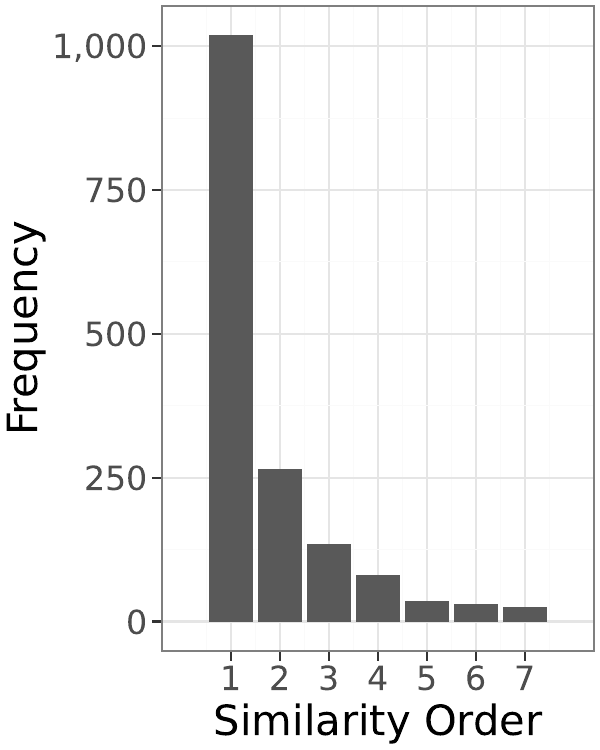}
  \end{subfigure}
  \begin{subfigure}{0.32\linewidth}
   \includegraphics[width=\textwidth]{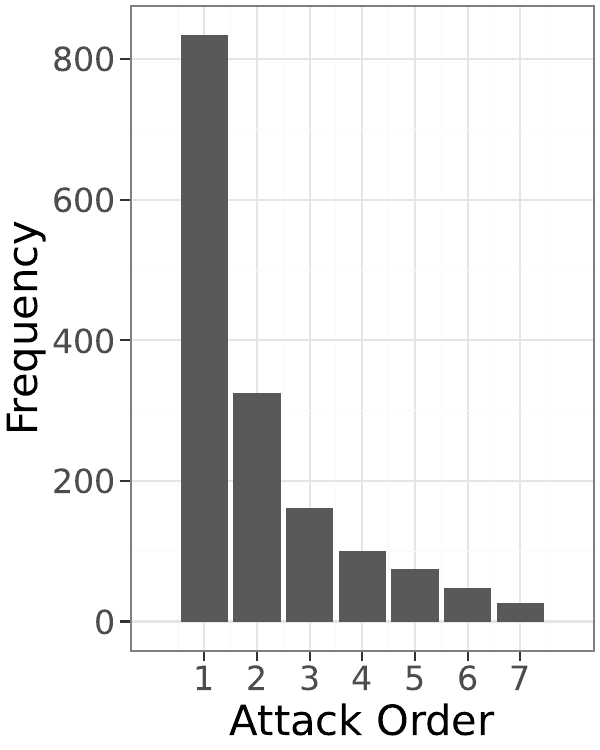}
  \end{subfigure}
  \caption{
    \label{fig:hist_competitions} 
    Histogram for the number of lookalike addresses per victim (left), the ranking of the $L_{\text{win}}$ in terms of similarity (middle), and timing (right).}
\end{figure}

\subsection{Benefit-cost analysis for attack groups}
\label{subsec:cost_benefit_analysis}
We next evaluate the attack profitability at the group level.
Table~\ref{tab:attack_group_profit} illustrates 1) the number of
successful attacks, 2) the total revenue, 3) the total cost, and 4) the
total profit for the top eight attack groups.
The revenue is the total amount of payoff transfers (i.e., phished
amount) in USD. 
The cost is 1) the transferred
value from the tiny transfer attacks (sent from $L$ to $V$) and 2) the
transaction fees for all poisoning transfers.
We defer the discussion of the address
generation costs to \S\ref{subsec:simulating_addr_generation}, and focus
here on transaction-related costs.

To accurately reflect the price fluctuation, We use the ETH-USD conversion rates from the CoinGecko API
for the day of the transaction. The profit is the difference between the
revenue and the cost. To highlight attack profitability for all attack
groups, Figure~\ref{fig:attack_group_profitability} illustrates the
relationship between costs ($x$-axis) and revenues ($y$-axis) where the
number is the group ID; the graph is in log-log
scale. We only include attack groups with at least one payoff transfer.
Revenues tend to
rise with cost, which is primarily dependent on the number of poisoning
transfers. 
The red line is the breakeven line---when costs and revenues
are the same. Most groups are above the line, and thus profitable, which 
indicates 
strong incentives to carry out blockchain address poisoning attacks.

\begin{table}[]
\caption{
  Attack group profitability
  \label{tab:attack_group_profit}
  }
  \centering
  \begin{adjustbox}{width=0.9\columnwidth, center}
    \begin{tabular}{@{}ccrrr@{}}
      \toprule
      \multirow{2}{*}{\textbf{Group}} & \multicolumn{1}{c}{\textbf{Nr. Success.}} & \multicolumn{1}{c}{\textbf{Revenue}}    & \multicolumn{1}{c}{\textbf{Total Cost}} & \multicolumn{1}{c}{\textbf{Profit}}     \\ 
      & \multicolumn{1}{c}{\textbf{Attacks}} & \multicolumn{1}{c}{(USD)} & \multicolumn{1}{c}{(USD)} & \multicolumn{1}{c}{(USD)}\\\midrule
      1     & 440         & 12,494,248 & 8,368,313  & 4,125,935  \\
      2     & 363         & 29,043,402 & 2,704,455  & 26,338,947 \\
      3     & 277         & 9,415,636  & 2,546,143  & 6,869,493  \\
      4     & 109         & 1,187,927  & 504,628    & 683,299    \\
      5     & 48          & 4,837,606  & 202,868    & 4,634,738  \\
      6     & 125         & 9,023,888  & 682,016    & 8,341,871  \\
      7     & 156         & 4,572,499  & 226,418    & 4,346,081  \\ 
      8     & 7           & 103,417    & 68,123     & 35,294    \\ \bottomrule
    \end{tabular}
  \end{adjustbox}
\end{table}

\begin{figure}
  \begin{minipage}[]{0.48\linewidth}
    \centering
    \includegraphics[width=0.8\linewidth]{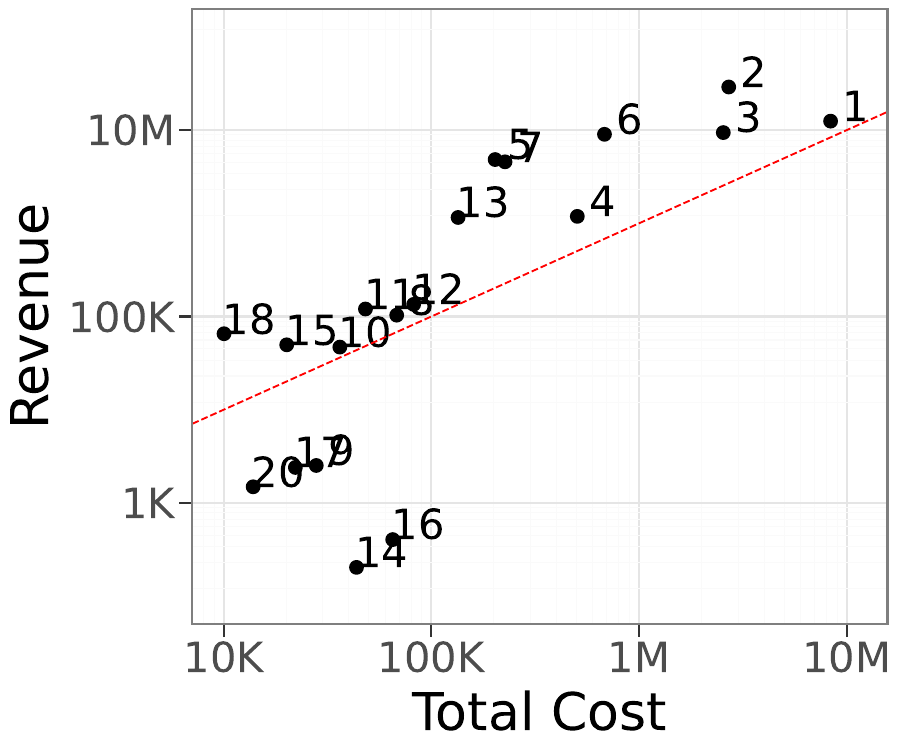}
    \captionof{figure}{Attack profit vs cost; red line: $y=x$}
    \label{fig:attack_group_profitability}
  \end{minipage}
  \hfill
  \begin{minipage}{0.48\linewidth}
  \centering
    \includegraphics[width=0.95\linewidth]{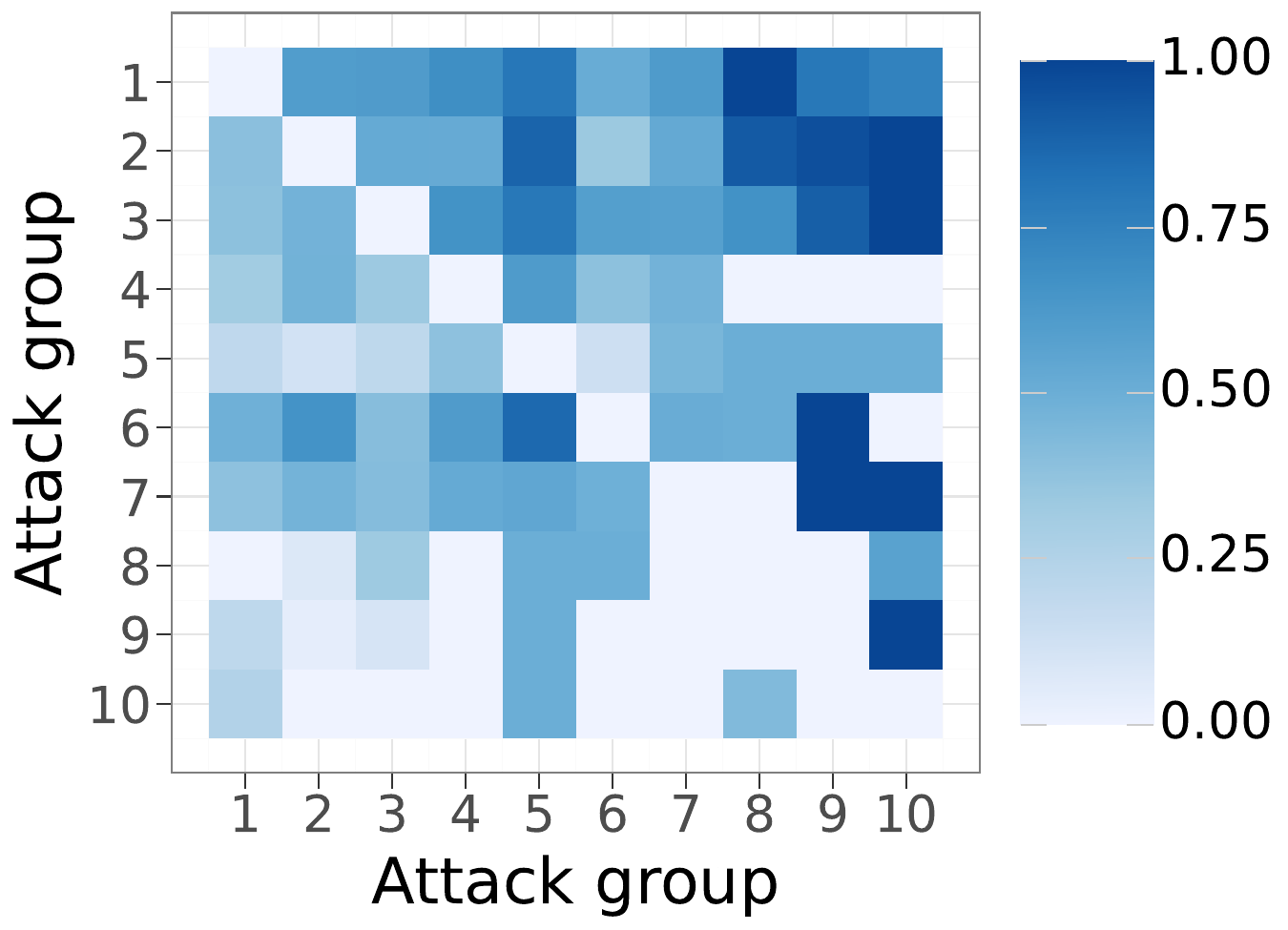}
    \captionof{figure}{Win-loss ratio between attack groups}
    \label{fig:win_loss_matrix}
  \end{minipage}
\end{figure}

In general, the top seven
groups are largely profitable. Despite the high variance in payoff amounts as shown in Table~\ref{tab:successful_attack_stats}, those groups manage to amortize
the revenue across many payoff transfers while some small groups do not.
Given the attack scale 
(i.e., Group 1 has invested over 8 million USD), some
groups appear to be well-funded cybercrime entities. However, none
of the attack instances comes from any sanctioned entities in the SDN
list~\cite{wahrstatter2023blockchain}.

We also investigate competition between groups. For each
successful attack, we collect the set of lookalike addresses
from each group and record which group wins over the others.
Figure~\ref{fig:win_loss_matrix} is the win-loss matrix; the
color indicates the win ratio for the group on the $y$-axis against the
group on the $x$-axis. The win-loss ratio of the top seven groups hovers
around 50\% within themselves. On the other hand, these
top groups mostly win against the smaller attack groups (Groups 8--10),
probably because larger groups tend to generate more (or have more
similar) lookalike addresses and attack attempts.

In summary, while some attack strategies (e.g., targeting wealthy accounts, cross-chain attacks) are effective, others (e.g., imitating hot wallets, generating lookalike addresses with large $d$) are not always optimal, suggesting that the attack damage could have been even more severe.

\section{Lookalike address generation}
\label{sec:address_generation}
Figures~\ref{fig:fake_addr_similarity_dist} and
\ref{fig:attack_group_fake_addr_dist} illustrate distinct distributions
in lookalike address similarity between Group~1 and others, suggesting vast differences in 
computational capabilities. We next mathematically formulate the address
generation problem, and simulate and measure the number of address
per second (APS) of various hardware and software implementations.

\subsection{Mathematical formulation}
\label{subsec:address_generation_math}
We assume that the attacker wants to match the first $a$ and the last
$b$ digits of the intended address and maximize $a+b=d$, with no preference for $a$
over $b$. For instance, the attacker is indifferent between $(a,
b)=(3,6)$ or $(4, 5)$ as long as $(a \geq 3, b \geq 4)$ holds, per 
\S\ref{subsec:detection_algo}. The attacker has a set of target
addresses ($\mathcal{R}$) they want to imitate as discussed in \S\ref{subsec:targeted_population}. 

We assume that the probability of generating a given address follows
a uniform random distribution over the entire address space, and that
the number of target addresses $|\mathcal{R}|$ is much smaller than the
total address space ($16^{40}$).
If $|\mathcal{R}|=r$, the probability of generating a target address is
\begin{equation}
\label{eq:collision_prob}
p = 1 - \left(\frac{16^d - 1}{16^d}\right)^r \ . 
\end{equation}
Using a first-order Taylor approximation
yields $p \approx r/16^d$.

The probability of finding a collision after $k$ trials follows the geometric distribution
$P(X=k)= p\cdot(1-p)^{k-1}$,
and the expected number of trials for a collision is $E[X] = 1/p \approx 2^{4d} / r$.  
In short, the expected number of trials exponentially increases with the the number of matched digits, 
and linearly decreases with the size of the target address space.

%
\subsection{Simulating address generation}
\label{subsec:simulating_addr_generation}
We implement and run, on various combinations of hardware and software,
a brute-forcing process to simulate lookalike address generation.
This allows us to measure the number of addresses per second
(APS) that can be generated, which in turn gives us a lower bound on the
computational resources attackers must possess to produce the lookalike
addresses we observed in the wild.

For address generation software, we have tested several combinations of libraries: 
\textit{web3py}\footnote{\url{https://github.com/ethereum/web3.py}.}, 
\textit{coincurve}\footnote{\url{https://ofek.dev/coincurve/}, a set of Python bindings for a heavily optimized C library for elliptic curve operations.}, 
\textit{pycryptodome}\footnote{\url{https://github.com/Legrandin/pycryptodome/}, the default hashing package used by \textit{web3py}.},
with/without \textit{concurrency}, and \textit{multiprocessing}.
We also implement the address generation scripts in Go (with concurrency), which yields results similar to those obtained with Python. 
We additionally use two popular vanity address software: Vanity-ETH,\footnote{\url{https://vanity-eth.tk/}}
a Javascript-based web application, and
Profanity2,\footnote{\url{https://github.com/1inch/profanity2}} a
C++-based GPU implementation. We elect, for simplicity, not to check whether
the generated address is in the set of target addresses $\mathcal{R}$, 
which will produce an optimistic bound on performance, and thus a
conservative bound on the attackers' implied hardware capabilities. 

We test our implementations on two hardware configurations, a MacBook Air (OS Version 13.1) with an 8-core Apple M1 base chip, and a 
server workstation (Ubuntu 22.04 LTS) with an Intel(R) Xeon(R) Silver 4214 CPU and an Nvidia RTX 8000 GPU.
Other hardware settings (e.g., different generations of Mac chips) only result in minor variations in outcomes. 
We run our simulations on the CPU on both machines and on the GPU on the second machine.
We summarize our simulation results in Table~\ref{tab:generating_addr}.
We focus on 1) a naive implementation which models what might be an expected ``first-attempt'' by an attacker, using 
the Python \textit{web3py} library, and 2) the best performing implementation: \textit{coincurve} + \textit{pycryptodome} + \textit{multiprocessing}.

The naive software implementation produces in the order of $10^{3}$--$10^4$ addresses per
second.
The more optimized software implementation produces an order of magnitude improvement ($10^4$--$10^5$).
The Vanity-ETH implementation is much slower than the standalone clients.
The CUDA implementation from Profanity2 is three orders of magnitude faster than our best CPU implementation with roughly half a billion APS. The Profanity2 developers self-report about a billion APS on similarly priced consumer hardware\footnote{\url{https://github.com/1inch/profanity2?tab=readme-ov-file\#benchmarks---current-version}}. 

\begin{table}[]
\centering
\caption{Generated addresses per second (APS)
\label{tab:generating_addr}}
\begin{adjustbox}{width=\columnwidth,center}
\begin{tabular}{@{}lrrr@{}}
\toprule
 & \textbf{Mac M1} & \textbf{Server (CPU-only)} & \textbf{Server (GPU)} \\ \midrule
Python (naive)     & 12,152          & 5,769   & -- \\
Python (optimized) & 81,660          & 460,665 & -- \\
Vanity-ETH         & $\approx$ 4,800 & --       & -- \\
Profanity2         & --               & --       & 516,437,000 \\ \bottomrule
\end{tabular}
\end{adjustbox}
\end{table}

Based on the distribution of the matching digits $d$ for each attack group, we next attempt to infer the attacker's hardware capability.  
We define two units: CPU-day and GPU-day, the number of addresses attackers can generate in one day per machine with a CPU and a GPU, respectively. 
From the above benchmarks, 1 CPU-day is $3.98\times  10^{10}$ addresses and 
and 1 GPU-day is $4.46 \times 10^{13}$, respectively. 
In the process of producing a $d$-digit match, the attacker can generate many addresses matching $d'<d$ digits in $\mathcal{R}$.  
Therefore, estimating the amount of computation needed to generate the maximum match $d$ for each group suffices. 
As shown in Figure~\ref{fig:attack_group_fake_addr_dist}, Group~1 produces a maximum 20-digit match while other groups (Groups~2, 3, and 4) achieve 14-digit matches. 
We calculate the number of expected trials for $d=20$ and, based on Table~\ref{tab:attack_group_general}, $r=10^6$. We estimate that Group~1 have used $3.0 \times 10^{7}$ CPU-days or 27,093 GPU-days. 
Group~1 would have needed to run over 41,000 CPUs around the clock over our whole two years of measurements to generate their lookalike addresses. We thus suspect they perform a substantial amount of their computation via GPUs.
Other groups ($d=14$) only used 1.81~CPU-days, or $1.6 \times 10^{-3}$ GPU-days.

To get a rough idea of the economic cost of computation, we use the AWS pricing model for similar GPUs and CPUs\footnote{GPUs: \url{https://aws.amazon.com/ec2/instance-types/g4/} and CPUs: \url{https://aws.amazon.com/ec2/pricing/on-demand/}}. 
For GPU, an hourly rate (yearly plan) for \textit{g4dn.16xlarge} (NVIDIA T4 tensor) is \$2.612 per hour; \$62.69 for one GPU-day. 
For CPU, an hourly rate (on-demand) for \textit{r6i.4xlarge} is \$1.008; \$24.19 for one CPU-day. 
Using these numbers, we estimate that Group~1 would have paid (at most) 1.7M USD to generate a 20-digit match, which is below its 4M USD profit described in Table~\ref{tab:attack_group_profit}. 
Furthermore, 
these numbers are likely higher than the costs incurred by self-hosting machines. 
Notably, we do not consider specialized hardware for address generation (e.g., ASICs), which would further reduce the generation time/cost for Group~1. 
For other groups with $d=14$, CPU costs ($1.81 \times 24.19 \approx 43$~USD) are dominated by transaction costs. 


\section{Mitigations}
\label{sec:mitigation}
We next suggest a few countermeasures to mitigate the success of blockchain address poisoning. 

\noindent\textbf{Protocol-level mitigations}: 
One mitigation is to map human-readable strings to each address, 
like DNS for IP addresses. Some
entities already have adopted domain names for wallet addresses
through Blockchain Domain Name System (BNS). Human-readable
addresses would reduce address poisoning as well as accidental
transfers (Appendix~\ref{sec:appendix:detect_accidental_transfers})
and enhance usability. BNS, however, fosters other
security risks such as typosquatting~\cite{muzammil2024typosquatting}, name hijacking (i.e., re-using
the expired domain names)~\cite{xia2022challenges}, or name
collisions~\cite{ito2024investigations}.

Another protocol-level solution is to raise the cost of generating an address and make the attack economically less effective~\cite{soska2021security}.
Adding 1 ms of latency per address results in (roughly) decreasing lookalike generation by 1,000 addresses per second. 
Verifiable delay functions~\cite{boneh2018verifiable} might be a good application for this purpose.
Alternatively, the chain could use a larger alphabet representation to
increase the cost of finding lookalikes---e.g., Bitcoin uses Base58, so
the cost of matching $d$ characters is $O(58^d)$ as opposed to $O(16^d)$ in
EVM-based chains. These additional costs could help render the attack
model economically impractical.

\noindent\textbf{Contract-level mitigations}: 
One solution is to disallow zero-value transfers in ``\texttt{transferFrom},'' despite the existence of use cases (e.g., testing by developers).
Specifically, we could theoretically modify the major stablecoin contracts (if they are upgradable) and mandate permission from the token sender, even when the transfer amount is zero. 
However, the community\footnote{\url{https://github.com/OpenZeppelin/openzeppelin-contracts/issues/3931}} has concluded that no changes to these contracts should be made, as removing zero-value transfers (in \texttt{transferFrom}) would break existing code (e.g., ``\texttt{\_becomeImplementation}'' in \texttt{CDaiDelegate.sol} (Compound) that relies on such zero-value transfers.)

Another potential (post-attack) solution is to quickly blacklist attack addresses. 
For the largest payoff transfer we observe, the Tether team (USDT) blocked the attacker's address within an hour, which prevented the attacker from moving the retrieved funds to another address.\footnote{Transaction blocking $L$: \texttt{0x2675\-aaf5\-db84\-f9f9\-66c4\-31ff\-82f2\-173d\-3251\-020b\-11b4\-939b\-4961\-ebed\-f92c\-78dd}.}

\noindent\textbf{Wallet-level mitigation}: 
Blockchain wallet or chain scanning services can improve usability.
For example, one can show more of the address 
to make lookalike addresses
easier to spot. 
One can also hide suspected poisoning
transfers (e.g., using our detection algorithms) from the account history
to prevent users from mistakenly selecting a lookalike; or at
least require confirmation when sending money to a lookalike
address.

\noindent\textbf{User-level mitigations}: 
Finally, users are the last line of defense. They could, for instance,
be trained to engage in safer practices, e.g., building an allow-list of
trusted addresses, or 
installing third-party extensions to flag phishing
addresses~\cite{yu2024don}. This, however, adds another layer of trust,
relies on the extension's effectiveness, and remains potentially vulnerable to
anti-detection strategies~\cite{oest2020phishtime}.

\section{Related work}
\label{sec:related_work}
The surge in blockchain popularity exposes users to a range of
cryptocurrency scams and online misconduct. Examples include
phishing websites~\cite{he2023txphishscope}, giveaway scams~\cite
{li2023double}, counterfeit tokens and rug pulls~\cite{gao2020tracking,
xia2021trade, ye2024interface}, market manipulation~\cite{xu2019anatomy,
li2021cryptocurrency, hamrick2018economics}, exchange/marketplace
scams~\cite{moore2013beware, tsuchiya2024identifying}, and Ponzi
schemes~\cite{vasek2018analyzing}. The above attacks exploit 1) the
absence of an intermediary~\cite{nakamoto2019bitcoin}, 2) pseudonymous
payments for scammers or criminals~\cite{christin2013traveling,
soska2015measuring}, 3) large price movements attracting inexperienced
investors~\cite{soska2021towards, kawai2023user} and inviting
mischief~\cite{tsuchiya2023misbehavior, li2023understanding}.

We highlight work particularly related to our discussion, with a focus on attack scale.
Gao et al.~\cite{gao2020tracking} identified 2,117 counterfeit tokens by applying keyword matching to the 100 most popular tokens. 
The authors investigated two techniques involving counterfeit tokens (airdrop and arbitrage), leading to total financial losses exceeding 17M USD over 7,000 victims. 
Xia et al.~\cite{xia2021trade} identified around 10,000 counterfeit tokens on Uniswap (50\% of all Uniswap tokens) through ground truth labels provided by the platform, address clustering, and machine learning-based techniques, leading to a 16M USD loss from approximately 
40,000 victims. 
He et al.~\cite{he2023txphishscope} introduced TxPhishScope, a tool designed to detect blockchain phishing websites. 
The authors identified 26,333 phishing websites with 3,486 phishing blockchain accounts.
Muzammil et al.~\cite{muzammil2024typosquatting} studied typosquatting on BNS, and identified over 26,000 typosquatting names and thousands of transactions sent to the addresses mapped to these squatting names. 
Other studies on free giveaway scams reported 24M--69.9M USD losses~\cite{li2023double}, including 872K USD losses from Twitter~\cite{li2023understanding}.
Based on these other works, blockchain address poisoning appears to be one of the largest (in terms of aggregate losses) and most widely targeted on-chain criminal activities.

Finally, we clarify our contributions in contrast to other contemporary
works, most notably Guan and Li~\cite{guan2024characterizing} and
Ye et al.~\cite{ye2024interface}. There are three major differences
between our approach and Guan and Li's. First, our clustering is
more fine-grained. While we use guilt-by-association techniques to
cluster attack instances (similar to theirs), crucially, we detect
copying bots, which will erroneously merge distinct groups into
one large cluster, as described in \S\ref{subsec:clustering_algo}
and Appendix~\ref{sec:appendix:copying_bots}. Second, we ran novel
experiments and analyses, including victim characteristics, success
conditions, group competitions, and address generation modeling, which
provide deeper insights into attacker strategies and computational
capability. Third, we uncovered 13 times more attacks (270M total) than
prior works and compared the results between the two chains.
Despite the similar chain structure, BSC requires more effort and
optimization due to the large number of transfers. (Our BSC database is
8.6 times larger than Ethereum’s). That dataset also allows us
to uncover cross-chain attacks.

Another contemporary piece, by Ye et al.~\cite{ye2024interface},
examines zero-value transfers, but not other types of poisoning
transfers, which we do cover. In Appendix~\ref{tab:comparison}, we
adjust our measurement period to provide a direct comparison of
detection outcomes between our work and those two studies.

While most industry reports are case studies, a Chainalysis’s blog
post~\cite{chainalysis2024anatomy} extended one case study (WBTC \$68M
loss) to identify an additional 82,031 lookalike addresses. While
commendable, this remains small compared to the whole ecosystem, as
shown by our work.

\section{Conclusion}
\label{sec:conclusion} 
We studied blockchain address poisoning, a form of phishing that exploits
visual similarities in hexadecimal wallet addresses
used in EVM-compatible chains (e.g., Ethereum, BSC).
We implemented a detection, identified large attack groups and their strategies, and inferred their capability through simulation. 
We showed blockchain
address poisoning presents a considerable threat, in total, 
270M attack attempts, 50M lookalike (malicious)
addresses, and, more chillingly, nearly 84M USD in monetary losses (from more than 6,600 cases) on Ethereum and BSC.
Worse, damages could be even higher if attackers optimized their
strategies. 

\section{Ethical considerations}
\label{sec:ethics}
While the concept of address poisoning is already the subject 
of many online discussions, we notified Etherscan and the MetaMask security team of our findings; MetaMask acknowledged that they are aware of address poisoning.  
Although attackers could potentially misuse some of our findings to improve
their strategies, the
main contribution of our paper is to demonstrate its practical severity in economic terms.
We hope that our work can help users recognize when they are in targeted populations, and accordingly, take precautionary measures. 
We also introduce possible protection strategies. 
Our proposed mitigations can hopefully inspire wallet operators. 

\section{Open science}
\label{sec:open_science}
We released the detection result dataset, including over 17 million attack attempts on Ethereum and successful payoff transfers.\footnote{\url{https://kilthub.cmu.edu/articles/dataset/Blockchain_Address_Poisoning_Companion_Dataset_/29212703/1}}
We also provide a Jupyter notebook explaining 1) how to access the dataset, 2) how to produce descriptive statistics such as the number of poisoning transfers, and 3) how to manually verify the payoff transfer on Etherscan (BSCscan).
This dataset will enable other researchers to validate our results as well as conduct further analysis. 
\section*{Acknowledgments}
This research was partially supported by a Sui Foundation Academic Grant, 
the Carnegie Mellon CyLab’s Secure Blockchain Initiative, 
and the Nakajima Foundation.

\appendix
\section{Investigating false negatives}
\label{sec:appendix:analyze_fn}
This appendix provides more details about the procedure we use to
categorize false negatives in \S\ref{subsec:detection_eval}. We
first obtain the set of victims $V$ that are seemingly attacked by
lookalike addresses $L$ that we had missed in the Guan and Li's dataset~\cite{guan2024characterizing} we used for evaluation. We then retrieve
each victim's ERC-20 or native ETH transfers on Etherscan. We follow the
flow chart laid out in Figure~\ref{fig:analyze_fn} and use the criteria
defined in the green boxes to determine the reason (in the blue boxes)
why we missed the lookalike address. ``Mislabeling'' indicates that
the false negatives were actually true negatives -- and that the issue
lied with the dataset we compared to. (There are 45 such cases.) The
asterisk~(*) indicates that we lower the similarity threshold if we
do not find the intended address $R$ with the current threshold $(3,
4)$. The ``unknown'' category (orange box) signifies that the reason
is (primarily) due to data collection quirks.
For example, we fail to capture the original transfer between the victim $V$ and the intended address $R$ if the transfer is outside of our measurement period (2 years) or uses \texttt{transferFrom} instead of \texttt{transfer}. 
(We capture \texttt{transferFrom} when used for poisoning transfers.)
These concern only 38 false negatives out of 164.

\begin{figure*}
  \centering
  \includegraphics[width=1\linewidth]{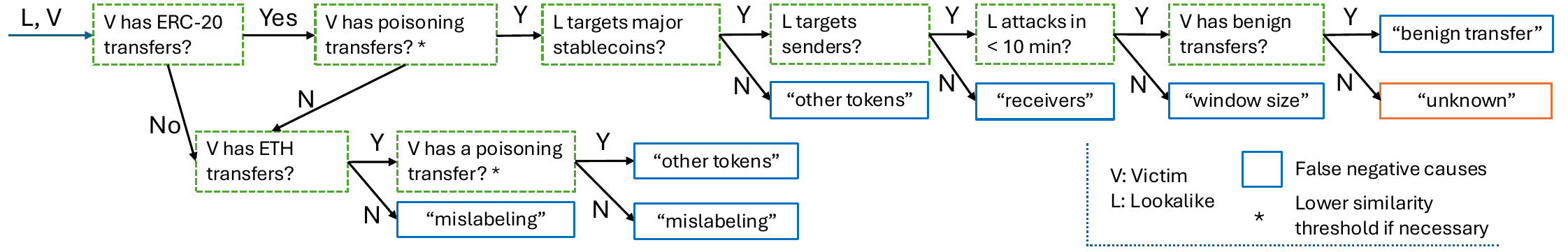}
  \caption{Flow chart of the identification of false negatives causes.}
  \label{fig:analyze_fn}
\end{figure*}

\section{Additional comparisons with other works}
\label{sec:appendix:comparison}
This appendix provides additional details on the performance comparison of 
our detection algorithm to
Ye et
al~\cite{ye2024interface} and Guan and Li~\cite{guan2024characterizing}.
For a fair comparison, here, we restrict the data we collected to
the same time period as those studies.

We detect more payoff transfers than Ye et al. We find 1,059
payoff transfers with 36.6M USD in losses while they found 560 transfers 
with \textit{at most} 28M USD in losses, including two non-address poisoning attacks they investigate. For poisoning
transfers, we identify 4.9M zero-value transfers, compared to
21M for Ye et al. It is unclear whether the difference comes from 1)
false positives on Ye et al.'s side (i.e., benign zero-value transfers), or 2) false negatives on our side.

Table~\ref{tab:comparison} summarizes our comparison with Guan and
Li~\cite{guan2024characterizing}. First, we find a similar total amount
of payoff transfers, corroborating the validity of both approaches.
Second, both approaches estimate similar attack costs in ETH, but not
in USD. Guan and Li use a fixed ETH-USD conversion rate (around
3,200~USD) whereas we query the price at the time of attack. Because
the exchange rate fluctuated between 1000 and 2000~USD for the
majority of the observation period, Guan and Li overestimate
attack costs by over 10~million USD, which results in a difference of
a 300\% on the return-on-investment (RoI). Third,
we have more than 2.2 million counterfeit token transfers and 3,500
counterfeit tokens. Guan and Li appear to only report counterfeit tokens
whose symbols are USDT or USDC. For example, out of
5,838 counterfeit tokens we find, only 1,922 use the exact same token symbols
whereas at least 740 tokens use the different encodings of USDT or USDC
(e.g., using a Cyrillic ``S'' and/or ``T'' in ``USDT''). Finally, our
technique misses attacks where $L$ matches $R$ in less than 7 digits.
Guan and Li do consider these attacks, which, however, are shown to have
a negligible impact in \S\ref{subsec:detection_eval} (and which come at
the increased risk of false positives).

\begin{table}[h]
    \caption{Comparison of detection performance
    \label{tab:comparison}}
    \begin{adjustbox}{width=\columnwidth,center}
    \begin{tabular}{@{}llllllll@{}}
    \toprule
           & \textbf{Payoff}   & \textbf{Cost ETH} & \textbf{ROI} & \textbf{Tiny} & \textbf{Zero-value} & \textbf{Counterfeit} & \textbf{CT} \\
           & \textbf{transfer} & \textbf{(USD)}    & \textbf{}    & \textbf{}     & \textbf{}           & \textbf{token}       & \textbf{}     \\ \midrule
Guan and Li~\cite{guan2024characterizing} & 76.79M            & 7918.2 (25.5M)    & 2.2          & 277.3K        & 7.85M               & 6.64M                & 2,333          \\
    This paper & 76.81M            & 8202 (14.9M)      & 5.1          & 283.9K        & 7.05M               & 8.87M                & 5,838          \\ \bottomrule
    \end{tabular}
    \end{adjustbox}
\end{table}

\section{Impact of the Birthday Paradox}
\label{sec:appendix:bday_thresh}
This section investigates whether our detection system mislabels poisoning transfers when the victim interacts with two similar-looking addresses by coincidence (i.e., due to the ``birthday paradox''). 
We define the number of accounts the victim interacts with as $r=|\mathcal{R}|$ and the likelihood of encountering, among these accounts, by pure chance, two addresses with the first three and the last four characters matching as $p$, a ``collision probability.''
By the birthday paradox, we can mathematically relate those two variables with: 
$p \approx 1-e^{{-r^2}/{2\times 16^{(3+4)}}}$. 
For instance, when the victim interacts with $r=$19,290 accounts, the probability $p$ becomes 0.5.

We examine whether our system should exclude a set of victims that present a collision probability greater than a given threshold $a$, i.e., $p\geq a$. We denote this set of victims by $\mathcal{V}_{p\geq a}$.
Figure~\ref{fig:bday_threshold} shows the number of victims ($|\mathcal{V}_{p>a}|$) that present a collision probability greater than $a$, on the $x$-axis.
Even at a small threshold $a=0.01$, we only find 664 accounts (0.04\% of total victims).
Excluding $\mathcal{V}_{p\geq a}$ does not seem to impact our detection performance. 
However, excluding these accounts results in numerous false negatives because attackers purposefully target very active accounts, as shown in \S\ref{subsec:targeted_population}. 
Figure~\ref{fig:bday_threshold} illustrates the number of attacks we would report if we excluded attacks on $\mathcal{V}_{p>a}$. 
For example, we would discover 16.2M million poisoning transfers at $a=0.01$, which is 1 million fewer than at $a=0.99$. 
Indeed, we find that most of those detected attacks are indeed poisoning transfers.
99.7\% of attacks targeting $\mathcal{V}_{p>0.99}$ are launched by the same attack accounts $\mathcal{A}$ that have also targeted victims $\mathcal{V}_{p < 0.01}$ (i.e., the population that is hardly susceptible to a collision).

False positives on payoff transfers are even more unlikely as they require: 1) the victim interacts with two similar addresses, 2) one of similar addresses unintentionally performs the benign transfer (e.g., a zero transfer for testing), and 3) the intended, the poisoning, and the payoff transfers happen in that correct order. 
We indeed do not find any false positives for the top 30 cases for both chains in \S\ref{subsec:detection_eval}. 

Those results imply that our system is robust against spurious collisions and that setting a low threshold reduces the performance of our detection system.
Hence, we decide to set $a=0.999$ to exclude victims that are interacting with a large enough number of addresses to be extremely susceptible to collisions, while capturing nearly all attacks. Setting such a high threshold only excludes 64 Ethereum accounts.

\begin{figure}
    \centering
    \begin{subfigure}{0.45\linewidth}
        \includegraphics[width=\textwidth]{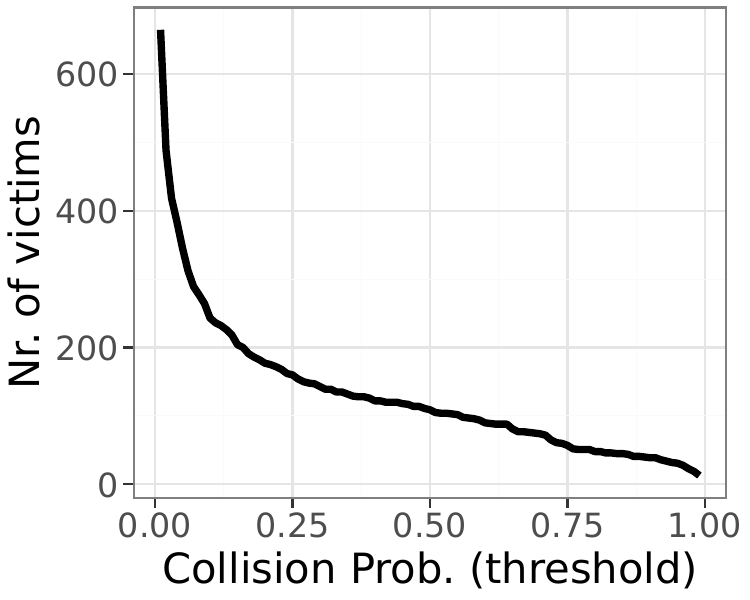}
    \end{subfigure}
    \begin{subfigure}{0.45\linewidth}
        \includegraphics[width=\textwidth]{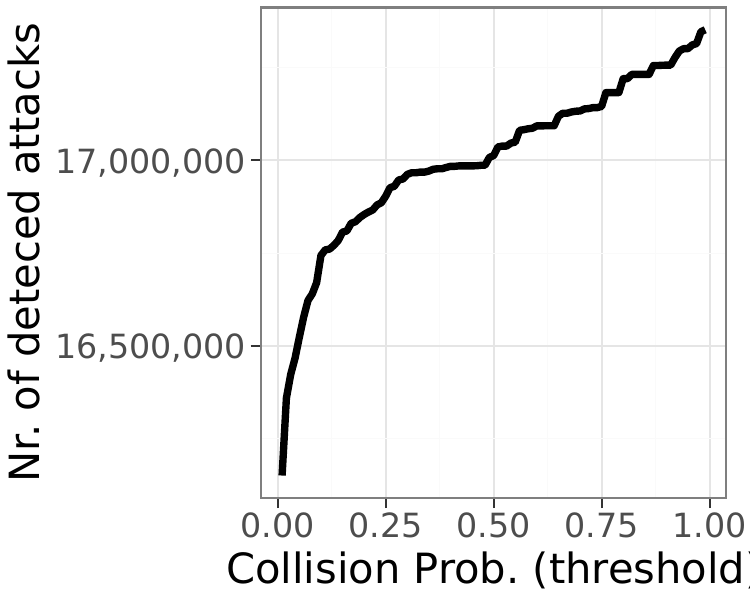}
    \end{subfigure}
    \caption{Left: Nr. of victims that are susceptible to a collision ($y$-axis) given a threshold ($x$-axis). Right: Nr. of attacks detected ($y$-axis) when we exclude those victims with a threshold ($x$-axis). The $x$-axis does not start with 0.}
    \label{fig:bday_threshold}
\end{figure}

\section{Accidental transfers}
\label{sec:appendix:detect_accidental_transfers}
Victims could accidentally send tokens to an incorrect, but not
malicious, destination address, due to a typing mistake. We focus
on those cases when two addresses (legitimate and lookalike) match
more than 20 (out of 40) characters, regardless of position. Given
current hardware capabilities, generating a lookalike address with
more than 20 matching digits is computationally unlikely (see
\S\ref{sec:address_generation} for details). Furthermore, if users
mistakenly specify an incorrect destination address, tokens are probably 
not recoverable, since it is very unlikely that anybody possesses the
private key of the mistyped address. We leverage this property, and
check whether one of the recipient addresses has ever sent any authentic
tokens;\footnote{To be more precise, we exclude zero-value transfers or
counterfeit token transfers which can be launched without private keys.}
if they have not, we hypothesize this transfer is an \emph{accidental
transfer} that we categorize separately from payoff transfers.

\noindent\textbf{Measurement results. } We identify 363 accidental transfers (i.e., not the result of a deliberate attack) from 295 victims and 336 mistyped destination addresses, totaling a 5.5M USD loss in Ethereum.
This high number suggests that some victims have sent assets to the same mistyped address multiple times, or made typing mistakes multiple times, over different intended addresses. 
We observe a similar number of accidental transfers in BSC: 318 cases from 206 victims and 241 mistyped addresses for a total of 57,562 USD lost. 
By manually looking at accidental transfers, we confirmed that those transfers appear to be typing mistakes, and noticed common human errors.
For instance, users 1) mistype a character with a neighboring key (i.e., low fat finger-distance~\cite{moore2010measuring}): ``4'' vs. ``3,'' ``3'' vs. ``e,'' 2) swap the order of two characters, or 3) mistype a character for something visually similar~\cite{szurdi2017email}: ``b'' vs ``d,'' ``e'' vs ``c,'' or ``6'' and ``b.'' 
To quantitatively assess typing mistakes, we calculate the Levenshtein distance $\ell$ (i.e., the number of operations necessary to replace a string with one another, including transposition) between two addresses. 
As expected, the distance is small; 84.95\% of typos have $\ell = 1$, and 96.69\% have $\ell \leq 2$ in Ethereum. 
We have similar results in BSC (86.48\% and 94.34\% for $\ell = 1$ and $\ell \leq 2$).

\section{Detecting copying bots}
\label{sec:appendix:copying_bots}
Here we explain why we use the attack ratio to exclude bots and show robustness in our results via temporal clustering.  

As discussed in \S\ref{subsec:clustering_algo}, we group two poisoning transfers if they 1) are in the same $TX$, 2) are launched by the same attacker $A$, or 3) specify the same lookalike address $L$. 
Our cluster considers two groups to be the same if they share even one link. 
We notice that two seemingly distinct large groups (based on their strategies and behaviors) are connected by a few attack addresses $A$ that behave differently from other attack addresses. 
We find that they appear to be the bots that simply copy transactions from multiple attack groups.
For instance, the account \texttt{0x7D575a7C732D1c502c07f18BB822D29CF7DBf9E8} copies not only address poisoning transactions\footnote{The original transaction: \texttt{0x725eaedf8857e243587020de97e-\\d503fb8bd8899bcbe8c685cf57fbce6810cc5}, \\the copy transaction: \texttt{0x5cc772d10b9e6fd61c7402967b5ae2ff6f-\\cae7a2ca9bef0e9c1c6202c2ea257c}} but also other transactions such as from MEV (Maximum Extractable Value) bots.\footnote{The original transaction: \texttt{0x0720fec6980c196b3545d273da2-\\b5b41987bf29524c6236cbf462d683df06057},\\ the copy transaction: \texttt{0x7aec84ca51fd5bdfcecf8717762-\\87b6732e8b8f27097e8d690c090cf6e6865bf}}
This type of copying behavior is relatively common in Ethereum or BSC, as discussed in previous work.\cite{qin2022quantifying, qin2023blockchain}

Those bots seem to copy the original transactions without properly understanding address poisoning. 
Without changing the lookalike address 
$L$, the payoff transfer goes to the original attacker regardless of who 
initiates the attack. These bots are helping other attackers.
We next explain why we use the attack ratio to detect copying bots over other methods (using the notion of time or removing all duplicated transactions).

\noindent\textbf{Using the notion of time. }
Copy transactions generally come after the original poisoning transactions (in terms of the block number), so removing the second or later transactions could be effective.
However, we notice that some copy transactions end up in the same block as the original transaction.   
This can be achieved by monitoring unconfirmed transactions in the peer-to-peer (P2P) network.  
For instance, the account \texttt{0xfffdcF2B3419C243A5eba5f051A64ad629362c9a} manages to include copy transactions in the same block as the 
original transaction. 
Based on the public mempool data from Flashbots,\footnote{\url{https://mempool-dumpster.flashbots.net/index.html}} we realize that this account's transactions do not appear in the public P2P network, while the original transactions do. 
This implies that some bots monitor the original transactions in the P2P layer, and directly submit their transactions to the block builders (to front-run the original transaction). 
As a result, we cannot systematically determine whether the first transaction is the original or the copy.

\noindent\textbf{Removing all duplicated transactions or transfers. }
While differentiating the original from the copy transactions remains impractical, we can simply remove all transactions involved in copying behavior (including the original).
If the bot simply copies the entire transaction, the data field in the transaction stays the same. 
While this method captures most of the copy transactions, we observe that some bots appear to slightly modify one of the parameters to potentially avoid detection, yielding a different data field. 

We can also exclude duplicated \textit{transfers} that have the same ($L$, $V$, $CT$ (or $AC$), and $v$ (value)).
This method appears to capture all the transactions from the bots.
However, we realize that attackers (not the copying bots) also resend their poisoning transfers several times against the same victim. 
We would remove more than 11 million (62\%) ``duplicated'' transfers, which significantly undermines the final clustering results (i.e., the size of the clusters becomes exceedingly small). 

\noindent\textbf{Removing potential bots by attack ratio. }
Only using attack transactions either leads to 1) the removal of too many non-bot transactions or 2) the inclusion of copying bot transactions. 
We decide to retrieve additional information of attack accounts $A$, and exclude $A$ with a low attack ratio---the ratio used for address poisoning, as introduced in \S\ref{subsec:clustering_algo}. 

We aim to set the threshold as low as possible while removing all the copying bots. 
Figure~\ref{fig:cluster_threshold} (left) illustrates the top 10 clusters when we vary the attack ratio threshold from 0 (removing no transfer) to 1 (removing all transfers).
The line indicates the same cluster.
We manually identify that all the erroneous merges we have spotted disappear with a threshold of 0.5; the largest cluster at threshold 0 breaks down into five different groups, which, as we show in \S\ref{subsec:clustering_results}, exhibit distinct behaviors. 
Figure~\ref{fig:cluster_threshold} (right) shows the ratio of removed transfers over the total number of transfers ($y$-axis) for each threshold value ($x$-axis).
Even setting a high threshold does not exclude many attacks, because attackers generally use $A$ only for address poisoning.
The attack addresses $\mathcal{A}$ with a threshold below 0.5 account for only 1.45\% of all transfers (represented by the vertical line on Figure~\ref{fig:cluster_threshold} (right)), which helps avoid removing too many attack instances and preserve the structure of the clusters. 
Groups remain stable up to a threshold of 0.7, but the group structure starts to collapse with a threshold above 0.8 (i.e., removing 6.64\% of all transfers).
Therefore, we choose a threshold of 0.5.

\begin{figure}
  \begin{subfigure}{0.55\linewidth}
    \includegraphics[width=\textwidth]{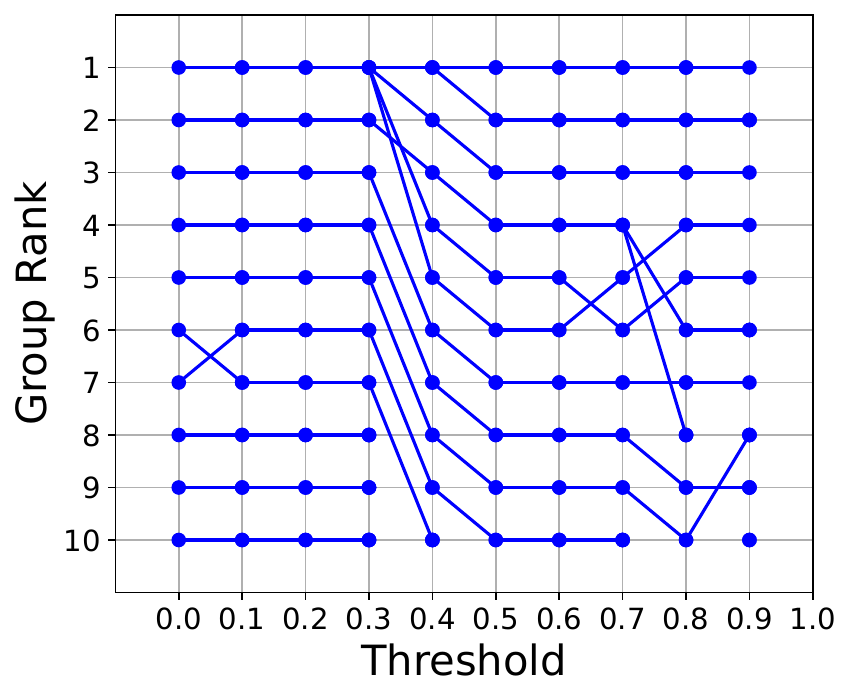}
  \end{subfigure}
  \begin{subfigure}{0.4\linewidth}
   \includegraphics[width=\textwidth]{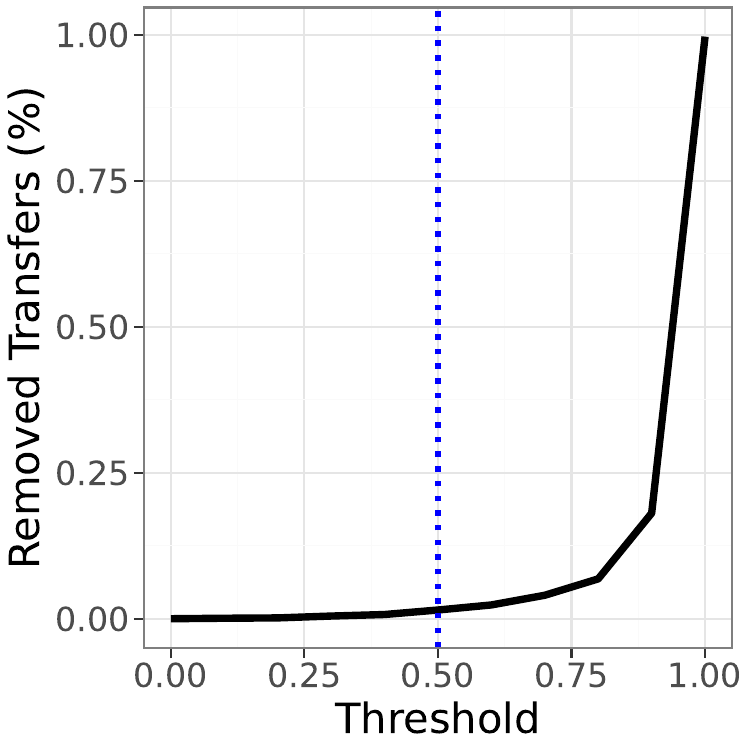}
  \end{subfigure}
  \caption{Left: Top 10 clusters ($y$-axis) based on the threshold ($x$-axis). Right: the ratio of removed transfers over the total number of transfers based on the threshold.}
  \label{fig:cluster_threshold} 
\end{figure}

To additionally validate our clustering robustness, we
perform a temporal clustering. We apply our algorithm for the
data until time $x$ and gradually increase $x$ over time to observe the
change in the results. Figure~\ref{fig:temporal_clustering}
shows how the top 10 clusters (based on the number of lookalike
addresses) evolve over time where the link indicates the same attack
group. The red line denotes Group~1; the blue lines represent other
groups. The left figure shows results with an attack ratio threshold of 0.5; the clusters appear to be stable over time. The right figure shows results
without any threshold (i.e., we do not exclude bots); large
clusters remain separate for a while, but eventually get connected
around Jul. and Aug. 2023, forming a single large cluster due to the
emergence of copying bots.

\begin{figure}[t]
    \centering
    \begin{subfigure}{0.47\linewidth}
        \includegraphics[width=\textwidth]{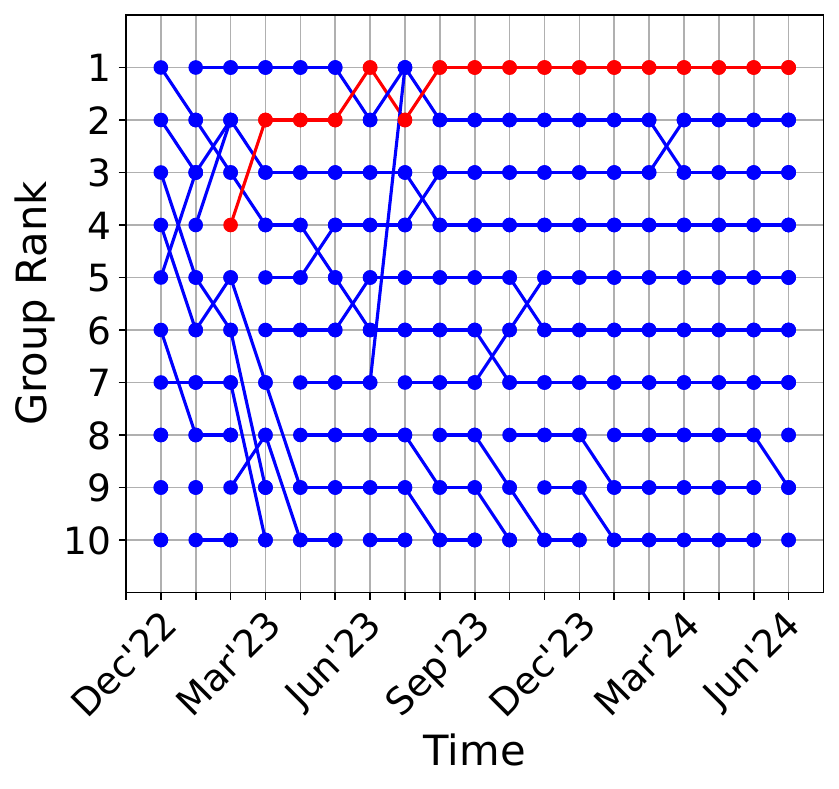}
        \caption{With a threshold of 0.5}
      \end{subfigure}
      \begin{subfigure}{0.47\linewidth}
       \includegraphics[width=\textwidth]{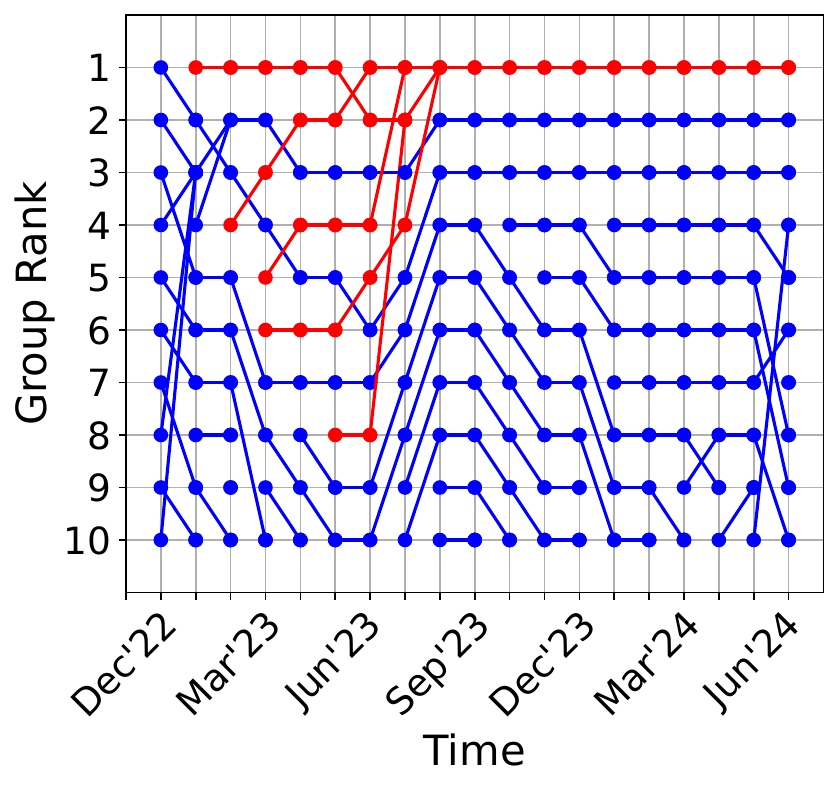}
       \caption{Without a threshold}
      \end{subfigure}
    \caption{Temporal clustering results. $x$-axis is time, $y$-axis is cluster rank of at time~$x$. Red: Group 1; blue: other groups. Links are used to denote the same group.}
    \label{fig:temporal_clustering}
\end{figure}

\section{Intended addresses}
\label{sec:appendix:intended_addr}
In this section, we focus on the intended addresses $R$---the victim's recipient, which attackers try to impersonate. 
Table~\ref{tab:intended_addr} lists the top ten intended addresses that the attackers impersonate the most (based on the number of lookalike addresses $L$ targeting each $R$). 
We label each address based on the third-party Arkham Intel website,\footnote{\url{https://intel.arkm.com/}} 
and aggregate the number of attacks targeting each $R$.
Table~\ref{tab:intended_addr} shows that many of $R$ are centralized exchange \textit{hot} addresses.
Since users do not directly interact with those hot wallets, we do not observe any successful cases for those addresses. 
Yet, those exchanges' deposit addresses frequently interact with those hot wallets, which may explain why the attackers' (algorithm) capture those intended addresses (as discussed in \S\ref{subsec:attack_strategies}).

\begin{table}[t]
    \centering
    \caption{Top 10 most targeted intended addresses
    \label{tab:intended_addr}}
    \begin{adjustbox}{width=\columnwidth,center}
    \begin{tabular}{@{}llrl@{}}
        \toprule
        Intended address $R$  & Label                & \multicolumn{1}{l}{Num of $L$} & Num of $TR$ \\ \midrule
        \texttt{0xA9D1e0...1d3E43}    & Coinbase Hot Wallet  & 143,764                        & 18,229                          \\
        \texttt{0x28C6c0...f21d60}    & Binance Hot Wallet   & 26,373                         & 16,147                          \\
        \texttt{0x974CaA...ECC400}    & Stake.com Hot Wallet & 3,254                          & 2,355                           \\
        \texttt{0xF7C8dA...38d921}    & Devout Deposit       & 5,242                          & 2,014                           \\
        \texttt{0xf89d7b...5EaA40}    & Bybit Hot Wallet     & 4,556                          & 1,702                           \\
        \texttt{0xBFCd86...23d6aD}    & HitBTC Hot Wallet    & 3,216                          & 1,569                           \\
        \texttt{0x731309...AB4D2a}    & BitGet Hot Wallet    & 7,547                          & 1,477                           \\
        \texttt{0x99870D...9AF1e2}    & HitBTC Deposit       & 9,700                          & 1,449                           \\
        \texttt{0x3CC936...aeCF18}    & MEXC Hot Wallet      & 6,603                          & 1,397                           \\
        \texttt{0x75e89d...1dcB88}    & MEXC Hot Wallet      & 1,629                          & 1,257                           \\ \bottomrule
    \end{tabular}
    \end{adjustbox}
\end{table}

\section{UI Design of wallets and chain scanners}
\label{sec:appendix:ui_design}
This section examines how address poisoning attacks appear in victims'
wallets or chain scanners. In particular, we visit those services in
Jan. 2025 and investigate how those services have reacted to the threat
of address poisoning.

To discover how victims see the attacks on their wallets and chain scanners, we first implement address poisoning transfers and attack our own accounts (i.e., attacking our accounts) on the Ethereum mainnet. 
While our attack uses minimal blockchain resources (three poisoning transfers), conducting our attack on the mainnet is critical because of the significant UI difference from the testnet. 
Specifically, we conduct our experiment on Dec. 20th, 2024, and use 159,364 gas for three poisoning transfers, which consumes about 1\% of one block (0.00014\% of gas usage for a day). 
We generate a lookalike address $L$ that matches the first six and the last four characters to $R$.
We follow to $AC$ and $CT$ implementations from the attackers who have uploaded their contracts' source codes on Etherscan and perform three poisoning transfers on our victim account $V$. 
Our attack can be found in $V=$\texttt{0xB6e84DF1cE401117C450221ccc6EF502cb0e2284}.  

First, we check the victim’s UI on blockchain wallets (as of Jan.     
2nd, 2025). Based on their wallet market share~\cite{qian2023most},     
we study MetaMask, Trust Wallet, and Phantom. All three services        
abbreviate addresses by only showing the first $a$ and last $b$         
characters. Metamask uses $(a,b) =(4,4)$, Trust Wallet uses (2, 4),     
and Phantom uses (4, 4). Trust Wallet and Phantom only display tiny     
transfers but hide zero-value or counterfeit token transfers. MetaMask  
does not seem to show any ERC-20 token transfer history in the first    
place. When initiating transfers, users no longer have options to       
select the recipient from ``recently used addresses'' but manually      
specify it on their own. The fact suggests that victims may rely more   
on blockchain scanning services to find the recipients.                 

Second, we check four major blockchain scanning services: Etherscan, Ethplorer, OkLink, and Arkham (as of Jan. 2nd, 2025).
The level of address abbreviation significantly differs between services.
Etherscan elides the middle part of the address while keeping the same number of characters at the beginning and at the end, while others significantly abbreviate one side (or do not abbreviate at all).
Since attackers mostly focus on collision on the first and last characters, displaying one side might be effective.
Next, we check the poisoning transfers. 
Most services (but Ethplorer) successfully hide zero-value transfers (by default) but fail to detect tiny transfers.
By changing settings, Etherscan and OkLink can show zero-value transfers, and Arkham displays counterfeit token transfers. 
It is generally easier for a service to flag zero-value transfers than tiny/counterfeit token transfers (by just observing their value); scanning services must implement more advanced heuristics to detect tiny or counterfeit token transfers\footnote{Etherscan successfully detects all poisoning transfers as of our visit on Jan. 13th, 2025.}. 
Hiding all zero-value transfers also hinders usability if users use zero-value transfers for (benign) testing purposes.
Most services show token icons or give a warning to $CT$ (based on their list of $AC$), which should be helpful to users.


\end{document}